\documentstyle[12pt,epsf,axodraw]{article}
\makeatletter
\def\appendix{\par\clearpage
  \setcounter{section}{0}
  \setcounter{subsection}{0}
  \@addtoreset{equation}{section}
  \def\@sectname{Appendix~}
  \def\theequation{\thesection.\arabic{equation}}
  \def\thesection{\Alph{section}}}
\makeatother

\renewcommand{\theequation}{\thesection.\arabic{equation}}

\begin{document}
\begin{titlepage}
\hskip 11cm \vbox{ \hbox{BUDKERINP/2001-33} \hbox{DFCAL-TH 01/2}
\hbox{June 2001}}

\vskip 0.3cm \centerline{\bf PHOTON-REGGEON INTERACTION VERTICES
IN THE NLA$^{~\ast}$}

\vskip 0.8cm \centerline{  V.S. Fadin$^{a, b~\dagger}$, D.Yu.
Ivanov$^{c, e~\ddagger}$ and M.I. Kotsky$^{a, d~\dagger\dagger}$}
\vskip .3cm \centerline{\sl $^a$ Budker Institute for Nuclear
Physics, 630090 Novosibirsk, Russia} \centerline{\sl $^b$
Novosibirsk State University, 630090 Novosibirsk, Russia}
\centerline{\sl $^c$ Institute of Mathematics, 630090
Novosibirsk, Russia} \centerline{\sl $^d$ Istituto Nazionale di
Fisica Nucleare, Gruppo collegato} \centerline{\sl di Cosenza,
Arcavacata di Rende, I-87036 Cosenza, Italy}
\centerline{\sl $^e$ Regensburg University, Germany}
\vskip 1cm

\begin{abstract}
We calculate  the effective vertices for the quark-antiquark  and
the quark-antiquark-gluon production in the virtual photon -
Reggeized gluon interaction. The last vertex is considered   at
the Born level; for the first one the one-loop corrections are
obtained. These vertices have a number of applications; in
particular, they  are necessary for calculation of the virtual
photon impact factor in the next-to-leading logarithmic
approximation.
\end{abstract}
\vfill
\hrule
\vskip.3cm
\noindent
$^{\ast}${\it Work supported in part by INTAS and in part by the
Russian Fund of Basic Researches.}
\vfill
$ \begin{array}{ll}
^{\dagger}\mbox{{\it e-mail address:}} &
 \mbox{FADIN@INP.NSK.SU}\\
\end{array}
$

$ \begin{array}{ll}
^{\ddagger}\mbox{{\it e-mail address:}} &
  \mbox{D-IVANOV@MATH.NSC.RU}
\end{array}
$

$ \begin{array}{ll} ^{\dagger\dagger}\mbox{{\it e-mail address:}}
&
 \mbox{KOTSKY@INP.NSK.SU}\\
\end{array}
$

\vfill
\vskip .1cm
\vfill
\end{titlepage}
\eject

\section{Introduction}
\setcounter{equation}{0}

Investigation of processes with  the Pomeron exchange remains to
be one of the important problems of high energy physics. A special
attention is attracted by the so called semi-hard processes,
where large values of typical momentum transfers $Q^2$ give a
possibility to use perturbative QCD for their theoretical
description. The most common basis for such description is given
by the  BFKL approach ~\cite{1}.  It became widely known since
discovery at HERA of the sharp rise of  the proton structure
function at decrease of the Bjorken variable $x$ (see, for
example,~\cite{2}). Recently the total cross section of the
interaction of two highly virtual  photons was measured at LEP.
This process, being the one-scale process, seems to be even more
natural for  the application  of the BFKL approach than the
two-scale process of the deep inelastic scattering at small $x$,
since here the evolution in $x$ described by the BFKL equation
does not interfere with the evolution in $Q^2$ described by the
DGLAP equation.

For a  consistent comparison with the experimental data the
theoretical predictions must be obtained in the next-to-leading
approximation (NLA), where  together with the leading terms
$(\alpha_s\ln(s))^n$  the terms
$\alpha_s(\alpha_s\ln(s))^n$ are also resumed. The radiative
corrections to the kernel of the  BFKL equation were calculated
several years ago~\cite{3}-\cite{8} and the explicit form of the
kernel of the equation in the NLA is known now~\cite{9,CC98} for
the case of forward scattering . But  the problem of calculation
in the NLA of the so called impact factors, which describe the
coupling of the Pomeron to the scattering particles, remains
unsolved.

Let us remind (see, for example, Ref. \cite{12} for the details),
that in the BFKL approach the relevant to the irreducible
representation ${\cal R}$ of the colour group in the $t$-channel
part  $\left({\cal A}_{\cal R} \right)_{AB}^{A^\prime B^\prime}$
of the scattering amplitude for the process $AB \rightarrow
A^\prime B^\prime$  at large c.m.s. energy $\sqrt{s} \rightarrow
\infty$ and fixed momentum transfer $q \approx q_\perp$ ($\perp$
means transverse to the initial particle momenta plane) is
expressed in terms of the Mellin transform of the Green function
of the two interacting Reggeized gluons $G_\omega^{(\cal R)}$ and
of the impact factors of the colliding particles $\Phi_{A^\prime
A}^{(\cal R, \nu)}$ and $\Phi_{B^\prime B}^{(\cal R, \nu)}$:
$$
{\cal I}m_s\left( {\cal A}_{\cal R} \right)_{AB}^{A^\prime B^\prime}
= \frac{s}{(2\pi)^{D-2}}\int\frac{d^{D-2}q_1}{\vec q_1^{~2}(\vec q_1
- \vec q)^2}\int\frac{d^{D-2}q_2}{\vec q_2^{~2}(\vec q_2
- \vec q)^2}
$$
\begin{equation}\label{11}
\times\sum_\nu\Phi_{A^\prime A}^{(\cal R, \nu)}(\vec q_1, \vec q, s_0)
\int_{\delta - i\infty}^{\delta + i\infty}\frac{d\omega}{2\pi i}\left[
\left( \frac{s}{s_0} \right)^\omega G_\omega^{(\cal R)}(\vec q_1, \vec q_2,
\vec q) \right]\Phi_{B^\prime B}^{(\cal R, \nu)}(-\vec q_2, -\vec q, s_0),
\end{equation}
where ${\cal I}m_s$ means the $s$-channel imaginary part,  the
vector sign is used for denotation of the transverse components,
$\nu$ enumerates the states in the representation $\cal R$,  $D =
4 + 2\epsilon$  is the space-time dimension different from $4$ to
regularize both infrared and ultraviolet divergencies and the
parameter $s_0$ is artificial and introduced for a convenience.
While the Green function obeys the generalized BFKL equation
\cite{12}
\begin{equation}\label{12}
\omega G_\omega^{(\cal R)}(\vec q_1, \vec q_2, \vec q) = \vec q_1^{~2}
(\vec q_1 - \vec q)^2\delta^{(D-2)}(\vec q_1 - \vec q_2) + \int\frac
{d^{D-2}k}{\vec k^{~2}(\vec k - \vec q)^2}{\cal K}^{(\cal R)}(\vec q_1,
\vec k, \vec q)G_\omega^{(\cal R)}(\vec k, \vec q_2, \vec q)
\end{equation}
with the NLA kernel ${\cal K}^{(\cal R)}$ and is completely
defined by this equation, the impact factors should be calculated
separately. The definition of  the NLA impact factors has been
given in Ref. \cite{12}; in the case of definite colours of $c$
and $c^\prime$ of the Reggeized gluons the  impact factor has a
form \cite{FFKP99}
$$ \Phi_{AA^\prime}^{cc^\prime}(\vec q_1, \vec
q, s_0) = \left( \frac{s_0}{\vec q_1^{~2}} \right)^{\frac{1}{2}
\omega(-\vec q_1^{~2})}\left( \frac{s_0}{(\vec q_1 - \vec q)^2}
\right)^{\frac{1}{2}\omega(-(\vec q_1 - \vec q) ^2)}
$$
$$
\times\sum_{\{f\}}\int\frac{d\kappa}{2\pi}\theta(s_\Lambda - \kappa)
d\rho_f\Gamma^c_{\{f\}A}\left( \Gamma^{c^\prime}_{\{f\}A^\prime} \right)^*
$$
\begin{equation}\label{13}
- \frac{1}{2}\int\frac{d^{D-2}k}{\vec k^{~2}(\vec k - \vec q)^2}
\Phi_{AA^\prime}^{c_1c_1^\prime(Born)}(\vec k, \vec q, s_0)\left(
{\cal K}_r^{Born} \right)_{c_1c}^{c_1^\prime c^\prime}(\vec k,
\vec q_1, \vec q)\ln\left( \frac{s_\Lambda^2}{s_0(\vec k - \vec
q_1)^2} \right),
\end{equation}
where  $\omega(t)$ is the Reggeized gluon trajectory and the
intermediate parameter $s_\Lambda$ should go to infinity. The
integration in the first term of the above equality is carried
out over the phase space $d\rho_f$ and over the squared invariant
mass $\kappa$ of the system $\{f\}$ produced in the fragmentation
region of the particle $A$, $\Gamma^c_{\{f\}A}$ are the
particle-Reggeon effective vertices for this production and the
sum is taken over all  systems $\{f\}$ which can be produced in
the NLA. The second term in  Eq. (\ref{13}) is the counterterm
for the LLA part of the first one, so that the logarithmic
dependence of both terms on the intermediate parameter $s_\Lambda
\rightarrow \infty$ disappears in their sum; ${\cal K}_r^{Born}$
is the part of the leading order  BFKL kernel  related to the real
gluon production (see Refs. \cite{FFKP99} for more details). It
was shown in Ref. \cite{FM99} that the definition (\ref{13})
guarantees infrared finiteness of the colourless particle impact
factors.

It is clear from above that for complete   NLA description  in the
BFKL approach one needs to know the impact factors, analogously as
in the DGLAP approach one should know not only the parton
distributions, but also the coefficient functions.

This paper is an extended version of the short note \cite{FIK},
which can be considered as the first step in the calculation of
the virtual photon impact factor in the NLA. We calculate here the
virtual photon-Reggeon effective vertices which enter  the
definition (\ref{13}) in the case when the particle $A$ is the
virtual photon. In the NLA the states which can be produced in the
Reggeon-virtual photon collision  are the quark-antiquark  and
the quark-antiquark-gluon ones. In the next Section we present the
effective  vertices for production of these states in the Born
approximation. This approximation is sufficient  to find in the
NLA the contribution to the virtual photon impact factor from the
quark-antiquark-gluon state. In the case of the quark-antiquark
state we need to know the effective production vertex with the
one-loop accuracy. Sections 3-5 are devoted to the calculation of
the one-loop corrections. In Sections 3  and 4 we consider the
two-gluon and the one-gluon exchange diagrams correspondingly; in
Section 5 the total one-loop correction is presented. The results
obtained are discussed in Section 6. Some details of the
calculation are given in Appendix A.

In the following  the photon-Reggeon effective vertices presented
in this paper will be used for the calculation of  the photon
impact factor. But they could have many other applications, for
example, in the diffractive production of quark jets and so on.

\section{The Born interaction vertices}
\setcounter{equation}{0}

In this section we present the vertices for the $q\bar q$ and the
$q\bar q g$ production in the Reggeon-virtual photon collision in
the Born approximation for the case of completely massless QCD.
These vertices can be obtained from the high energy amplitudes
with the octet colour state and the negative signature in the
$t$-channel for collision of the virtual photon with any
particle, if the corresponding system is produced in the virtual
photon fragmentation  region. For simplicity we always consider
collision of the virtual photon with the momentum $p_A$ and the
quark with the momentum $p_B$. We use everywhere below the
Feynman gauge for the gluon field, the Sudakov decomposition of
momenta
$$
p = \beta p_1 + \alpha p_2 + p_\perp,\ \ \ \alpha = \frac{p^2 +
\vec p^{~2}}{s\beta},\ \ \ p_1^2 = p_2^2 = 0,
$$
\begin{equation}\label{21}
s = 2p_1p_2 \rightarrow \infty,\ \ \  \vec p^{~2} \equiv -
p_\perp^2,
\end{equation}
with the lightcone basis in the longitudinal space  defined by
\begin{equation}\label{22}
p_A = p_1 - \frac{Q^2}{s}p_2,\ \ \ p_A^2 = - Q^2,\ \ \ p_B =
p_2,\ \ \ p_B^2 = 0~,
\end{equation}
and the usual trick of retaining only the first term in the
decomposition of the metric tensor
\begin{equation}
g^{\mu\nu} = \frac{2p_2^\mu p_1^\nu}{s}+\frac{2p_1^\mu
p_2^\nu}{s}+g_\perp^{\mu\nu} \rightarrow \frac{2p_2^\mu
p_1^\nu}{s} \label{z}
\end{equation}
in the numerator of the gluon propagator connecting vertices
$\mu$ and $\nu$ with momenta predominantly along $p_1$ and $p_2$
respectively.  The virtual photon polarization vector $e$ is
taken in the gauge $ep_2$=0, so that
\begin{equation}\label{23}
e = e_\perp + \frac{2ep_1}{s}p_2~.
\end{equation}
Then the  polarization vector $\tilde{e}$ in the usual gauge
$\tilde{e}p_A=0$ is
\begin{equation}\label{e}
\tilde{e} = e + \frac{ep_1}{Q^2}p_A~,
\end{equation}
so that in the case of the longitudinal polarization, when
$\tilde{e}_L^2=1$, we have $e_Lp_1=Q$.

Let us start with the calculation of the quark-antiquark
production vertex. The diagrams of the production process
contributing  in the Regge asymptotics are shown in Fig. 1.

\begin{figure}[tb]
\begin{center}
\begin{picture}(250,80)(0,-10)

\ArrowLine(0,0)(50,0) \ArrowLine(50,0)(100,0)
\ArrowLine(30,50)(100,70) \ArrowLine(100,30)(30,50)
\Photon(0,50)(27,50){3}{4} \ArrowLine(27,50)(30,50)
\Gluon(55,43)(55,3){3}{4} \ArrowLine(55,3)(55,0)

\Text(0,60)[l]{$p_A$} \Text(0,-10)[l]{$p_B$}
\Text(100,-10)[r]{$p_{B^\prime}$} \Text(65,20)[l]{$q$}
\Text(105,70)[l]{$k_1$} \Text(105,30)[l]{$-k_2$}

\ArrowLine(150,0)(200,0) \ArrowLine(200,0)(250,0)
\ArrowLine(180,50)(250,70) \ArrowLine(250,30)(180,50)
\Photon(150,50)(177,50){3}{4} \ArrowLine(177,50)(180,50)
\Gluon(205,58)(205,3){3}{6} \ArrowLine(205,3)(205,0)

\Text(150,60)[l]{$p_A$} \Text(150,-10)[l]{$p_B$}
\Text(250,-10)[r]{$p_{B^\prime}$} \Text(215,20)[l]{$q$}
\Text(255,70)[l]{$k_1$} \Text(255,30)[l]{$-k_2$}

\end{picture}
\end{center}
\caption[]{The lowest order Feynman diagrams for the process
$\gamma^* Q \rightarrow (q\bar q)Q$.}
\end{figure}
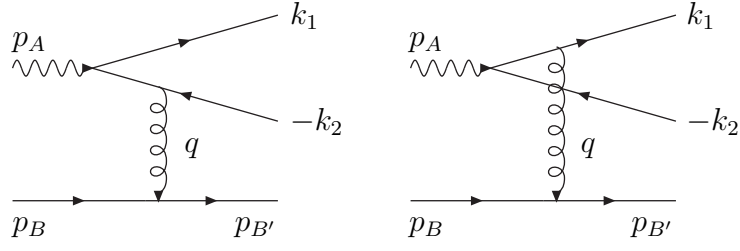

As was already mentioned, the quark-antiquark pair is produced in
the photon fragmentation region, so that the invariant mass
$\sqrt{\kappa}$ of the pair is of order of typical transverse
momenta and doesn't grow with $s$. The Regge form of the
production amplitude ${\cal A}_{Q\gamma^* \rightarrow Qq\bar
q}^{(0)}$ is
\begin{equation}\label{24}
{\cal A}_{Q\gamma^* \rightarrow Qq\bar q}^{(0)} = \Gamma_
{\gamma^*q\bar q}^{c(0)}\frac{2s}{t}\Gamma_{QQ}^{c(0)},
\end{equation}
where $t = q^2$ and $\Gamma_{\gamma^*q\bar q}^{c(0)}$ and
$\Gamma_{QQ}^{c(0)}$ are corresponding particle-Reggeon effective
vertices in the Born approximation. Let us note, that the
amplitude of Fig. 1 has automatically only the octet colour state
and the negative signature in the $t$-channel, so that it is not
necessary here to perform any projection. The notations for all
momenta are shown in Fig. 1. The quark-Reggeon vertex is known up
to the NLA accuracy and its Born part is
\begin{equation}\label{25}
\Gamma_{QQ}^{c(0)} = gt^c_{B^\prime B}\bar u_{B^\prime}\frac{\not p_1}{s}u_B,
\end{equation}
where $g$ is the coupling constant, $t^c$ are the colour group
generators in the fundamental representation and $u$ is a quark
spinor wave function. Then, comparing Eqs. (\ref{24}), (\ref{25})
with the explicit form of the amplitude given  by the diagrams of
Fig. 1, one can easily obtain for $\Gamma_{\gamma^*q\bar
q}^{c(0)}$ the diagrammatic representation of Fig. 2
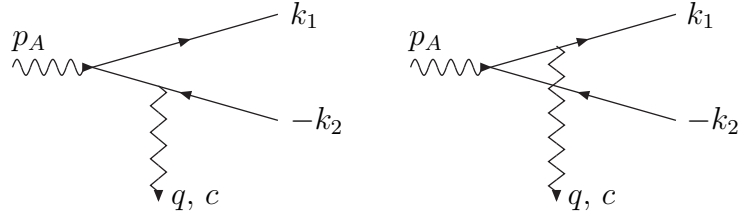
\begin{figure}[tb]
\begin{center}
\begin{picture}(250,80)(0,0)
\ArrowLine(30,50)(100,70) \ArrowLine(100,30)(30,50)
\Photon(0,50)(27,50){3}{4} \ArrowLine(27,50)(30,50)
\ZigZag(55,43)(55,3){3}{4} \ArrowLine(55,3)(55,0)

\Text(0,60)[l]{$p_A$} \Text(60,0)[l]{$q,\, c$}
\Text(105,70)[l]{$k_1$} \Text(105,30)[l]{$-k_2$}

\ArrowLine(180,50)(250,70) \ArrowLine(250,30)(180,50)
\Photon(150,50)(177,50){3}{4} \ArrowLine(177,50)(180,50)
\ZigZag(205,58)(205,3){3}{6} \ArrowLine(205,3)(205,0)

\Text(150,60)[l]{$p_A$} \Text(210,0)[l]{$q,\, c$}
\Text(255,70)[l]{$k_1$} \Text(255,30)[l]{$-k_2$}

\end{picture}
\end{center}
\caption[]{Schematic representation of the vertex
$\Gamma^{c(0)}_{\gamma^*q\bar q}$.}
\end{figure}
(see Ref. \cite{FM99}), where the zig-zag lines represent the
Reggeon with the momentum
\begin{equation}\label{26}
q = - \frac{\kappa + Q^2 + \vec q^{~2}}{s}p_2 + q_\perp,\ \ \ t =
q^2 = q_\perp^2 = - \vec q^{~2},
\end{equation}
and the colour index $c$. The lowest order effective vertices for
interaction of the Reggeon with quarks and gluons are defined in
Fig. 3 (see Ref. \cite{FM99}).

\begin{figure}[tb]
\begin{picture}(300,90)(0,-20)

\ArrowLine(10,80)(50,50) \ArrowLine(50,50)(90,80)
\ZigZag(50,50)(50,3){3}{4} \ArrowLine(50,3)(50,0)

\Text(55,0)[l]{$c$} \Text(100,40)[l]{$=igt^c\frac{\not p_2}{s}$.}
\Text(100,-20)[c]{$(a)$}

\Gluon(160,80)(196,53){3}{4} \Gluon(240,80)(204,53){3}{4}
\ArrowLine(196,53)(200,50) \ArrowLine(204,53)(200,50)
\ZigZag(200,50)(200,3){3}{4} \ArrowLine(200,3)(200,0)

\Text(205,0)[l]{$k_1+k_2,\, c$} \Text(170,80)[l]{$k_1,\, a,\,
\mu$} \Text(245,80)[l]{$k_2,\, b,\, \nu$}
\Text(250,40)[l]{$=-igT^c_{ab}\biggl[
g^{\mu\nu}\frac{p_2(k_2-k_1)}{s} +
\frac{p_2^\mu}{s}(2k_1+k_2)^\nu$} \Text(270,10)[l]{$-
\frac{p_2^\nu}{s}(2k_2+k_1)^\mu +
2\frac{(k_1+k_2)^2}{s}\frac{p_2^\mu p_2^\nu} {p_2(k_1-k_2)}
\biggr]$.} \Text(250,-20)[c]{$(b)$}

\end{picture}
\caption[]{The quark-quark-Reggeon and the gluon-gluon-Reggeon
effective vertices. $T^c$ is the colour group generator in the
adjoint representation.}
\end{figure}
The vertex $\Gamma_{\gamma^*q\bar q}^{c(0)}$ can be obtained from
the diagrams of Fig. 2 by the usual Feynman rules as the
amplitude of the quark-antiquark production in collision of the
virtual photon with the  Reggeon. This procedure gives us the
result
\begin{equation}\label{27}
\Gamma_{\gamma^*q\bar q}^{c(0)} = - eq_fgt^c_{i_1i_2}\bar u_1\left(
\frac{\hat \Gamma_1}{t_1} - \frac{\hat \Gamma_2}{t_2} \right)\frac
{\not p_2}{s}v_2 = - eq_fgt^c_{i_1i_2}\left( \left[ \bar u_1\frac
{\hat \Gamma_1}{t_1}\frac{\not p_2}{s}v_2 \right] - \left[
1 \leftrightarrow 2 \right] \right),
\end{equation}
where $eq_f$ is the electric charge of the produced quark, $i_1$
and $i_2$ are the  colour indices of the quark and antiquark
correspondingly, $v$ is the spinor wave function of the produced
antiquark,
$$
\hat\Gamma_1 = \frac{1}{x_1}\left( 2x_2(ek_1) - \not e_\perp{\not k_1}
_\perp \right),\ \ \ \hat\Gamma_2 = \frac{1}{x_2}\left( 2x_1(ek_2) -
{\not k_2}_\perp\not e_\perp \right),
$$
\begin{equation}\label{28}
t_i = (p_A - k_i)^2 = - \frac{\vec k_i^{~2} + x_1x_2Q^2}{x_i},
\end{equation}
with the variables $x_i$ defined by the Sudakov decompositions of
the produced quark and antiquark momenta
\begin{equation}\label{29}
k_i = x_ip_1 + \frac{\vec k_i^{~2}}{sx_i}p_2 + {k_i}_\perp,\, \,
k_i^2= 0~, \,\,i=1,2~.
\end{equation}
The substitution  $1 \leftrightarrow 2$ in Eq.
(\ref{27}) means replacement quark $\leftrightarrow$ antiquark ,
i.e. $x_1\leftrightarrow x_2~, \vec k_1\leftrightarrow \vec k_2$
together with replacement of the polarizations.  Validity of the
second equality in (\ref{27}) can be easily verified using the
charge conjugation matrix.  We will need later the Born effective
vertex $\Gamma_{\gamma^*q\bar q}^{c(0)}$  also in the helicity
representation for the case of the space-time dimension $D$ equal
$4$. To obtain it we use the polarization matrix
\begin{equation}\label{210}
\hat \rho \equiv (v_2\bar u_1) =
\frac{1}{\sqrt{x_1x_2}}\frac{1}{4}\left[ \left( x_2k_1 + x_1k_2
-\kappa\frac{p_2}{s} \right)^\mu - 2i\xi
e^{\mu\nu\sigma\rho}{k_2}_\nu{k_1}_\sigma\frac{{p_2}_\rho}{s}
\right]\gamma_\mu\left( 1 - \xi\gamma_5 \right),
\end{equation}
where
\begin{equation}
e^{0123}=1~, \,\,\,
\gamma_5=i\gamma^0\gamma^1\gamma^2\gamma^3~,\,\,\,
\end{equation}
and $\xi = \pm 1$  is a double helicity of the produced quark.
The polarization matrix  satisfies the  evident relations
\begin{equation}
\not k_2\hat\rho = \hat\rho\not k_1 =\left( 1 - \xi\gamma_5
\right)\hat\rho=\hat\rho\left( 1 +\xi\gamma_5 \right)  =0~.
\end{equation}
For the virtual photon polarization vector we also use the
helicity representation
\begin{equation}\label{211}
e^\mu(\lambda) = \frac{1}{\sqrt{-2t}}\left[ \left(
\delta_{\lambda, 1} + \delta_{\lambda, -1} \right)\left(
q_\perp^\mu + 2i\lambda
e^{\mu\nu\sigma\rho}q_\nu{p_1}_\sigma\frac{{p_2}_\rho}{s} \right)
+ \delta_{\lambda, 0} \sqrt{-2tQ^2}\frac{2p_2^\mu}{s} \right],\ \
\ \lambda = 0, \pm 1.
\end{equation}
Using Eqs. (\ref{27}) -  (\ref{211}) we get
$$
\Gamma_{\gamma^*q\bar q}^{c(0)} = -
\frac{2eq_fgt^c_{i_1i_2}}{\sqrt{-2tx_1x_2}}\left( \left[
\frac{x_2}{t_1} \left\{ \delta_{\lambda, 0}\sqrt{2}qQx_1x_2 -
\right. \right. \right.
$$
$$
\left. \left. \left. - (\vec k_1\vec q + i\lambda P)\left( x_2
\delta_{\lambda, -\xi} - x_1\delta_{\lambda, \xi} \right) \right\}
\right] - \left[ 1 \leftrightarrow 2 \right] \right) =
$$
$$
= \frac{2eq_fgt^c_{i_1i_2}}{\sqrt{-2tx_1x_2}}\left[
\delta_{\lambda, 0}\sqrt{2}qQx_1x_2\left( \frac{x_1}{t_2} -
\frac{x_2}{t_1} \right)\right.
$$
\begin{equation}\label{212}
\left.+\left( x_2\delta_{\lambda, -\xi}-x_1\delta_{\lambda,
\xi}\right)\left(\frac{x_2}{t_1}(\vec k_1\vec q + i\lambda P)
+\frac{x_1}{t_2}(\vec k_2\vec q -i\lambda P)\right) \right],
\end{equation}
where $q = |\vec q|$,
\begin{equation}\label{213}
P =
2e^{\mu\nu\sigma\rho}{k_1}_\mu{k_2}_\nu{p_1}_\sigma\frac{{p_2}_\rho}{s},
\end{equation}
with the property $P^2 = \vec k_1^{~2}\vec k_2^{~2} - (\vec
k_1\vec k_2)^2$, and the replacement $(1 \leftrightarrow 2)$  is
$x_1 \leftrightarrow x_2~,\, \vec k_1 \leftrightarrow \vec k_2,\
\xi \leftrightarrow -\xi$.

Next we do is the calculation of the quark-antiquark-gluon
production effective vertex $\Gamma_{\gamma^*q\bar qg}^{c(0)}$.
It can be obtained through the usual Feynman rules with the
elementary Reggeon vertices defined at Fig. 3 as the amplitude of
the quark-antiquark-gluon production in the virtual photon-Reggeon
collision represented by the diagrams of Fig. 4, where the
denotations of momenta are presented. The colour indices of the
Reggeon and the emitted gluon are $c$ and $b$ respectively.
\begin{figure}[tb]
\begin{picture}(430,130)(10,-100)

\ArrowLine(40,50)(110,70) \ArrowLine(110,30)(40,50)
\Photon(10,50)(37,50){3}{4} \ArrowLine(37,50)(40,50)
\ZigZag(65,43)(65,3){3}{4} \ArrowLine(65,3)(65,0)
\Gluon(65,57)(106.5,45){3}{4} \ArrowLine(106.5,45)(110,44)

\Text(10,60)[l]{$p_A$} \Text(70,0)[l]{$q$}
\Text(115,70)[l]{$k_1$} \Text(115,30)[l]{$-k_2$}
\Text(115,44)[l]{$k$}

\ArrowLine(190,50)(260,70) \ArrowLine(260,30)(190,50)
\Photon(160,50)(187,50){3}{4} \ArrowLine(187,50)(190,50)
\ZigZag(215,43)(215,3){3}{4} \ArrowLine(215,3)(215,0)
\Gluon(207.5,45)(254.75,54){3}{5} \ArrowLine(254.75,54)(260,55)

\Text(160,60)[l]{$p_A$} \Text(220,0)[l]{$q$}
\Text(265,70)[l]{$k_1$} \Text(265,30)[l]{$-k_2$}
\Text(265,55)[l]{$k$}

\ArrowLine(340,50)(410,70) \ArrowLine(410,30)(340,50)
\Photon(310,50)(337,50){3}{4} \ArrowLine(337,50)(340,50)
\ZigZag(365,43)(365,3){3}{4} \ArrowLine(365,3)(365,0)
\Gluon(383.75,37.5)(404.75,44){3}{2} \ArrowLine(404.75,44)(410,45)

\Text(310,60)[l]{$p_A$} \Text(370,0)[l]{$q$}
\Text(415,70)[l]{$k_1$} \Text(415,30)[l]{$-k_2$}
\Text(415,44)[l]{$k$}


\ArrowLine(110,-30)(40,-50) \ArrowLine(40,-50)(110,-70)
\Photon(10,-50)(37,-50){3}{4} \ArrowLine(37,-50)(40,-50)
\ZigZag(65,-57)(65,-97){3}{4} \ArrowLine(65,-97)(65,-100)
\Gluon(65,-43)(106.5,-55){3}{4} \ArrowLine(106.5,-55)(110,-56)

\Text(10,-40)[l]{$p_A$} \Text(70,-100)[l]{$q$}
\Text(115,-30)[l]{$-k_2$} \Text(115,-70)[l]{$k_1$}
\Text(115,-56)[l]{$k$}

\ArrowLine(260,-30)(190,-50) \ArrowLine(190,-50)(260,-70)
\Photon(160,-50)(187,-50){3}{4} \ArrowLine(187,-50)(190,-50)
\ZigZag(215,-57)(215,-97){3}{4} \ArrowLine(215,-97)(215,-100)
\Gluon(207.5,-55)(254.75,-46){3}{5}
\ArrowLine(254.75,-46)(260,-45)

\Text(160,-40)[l]{$p_A$} \Text(220,-100)[l]{$q$}
\Text(265,-30)[l]{$-k_2$} \Text(265,-70)[l]{$k_1$}
\Text(265,-45)[l]{$k$}

\ArrowLine(410,-30)(340,-50) \ArrowLine(340,-50)(410,-70)
\Photon(310,-50)(337,-50){3}{4} \ArrowLine(337,-50)(340,-50)
\ZigZag(365,-57)(365,-97){3}{4} \ArrowLine(365,-97)(365,-100)
\Gluon(383.75,-62.5)(404.75,-56){3}{2}
\ArrowLine(404.75,-56)(410,-55)

\Text(310,-40)[l]{$p_A$} \Text(370,-100)[l]{$q$}
\Text(415,-30)[l]{$-k_2$} \Text(415,-70)[l]{$k_1$}
\Text(415,-56)[l]{$k$}

\end{picture}
\begin{picture}(430,100)(10,-10)
\ArrowLine(40,50)(110,70) \ArrowLine(110,30)(40,50)
\Photon(10,50)(37,50){3}{4} \ArrowLine(37,50)(40,50)
\Gluon(65,43)(65,23){3}{2} \ZigZag(65,23)(65,3){3}{2}
\ArrowLine(65,3)(65,0) \Gluon(65,23)(106.5,11){3}{4}
\ArrowLine(106.5,11)(110,10)

\Text(10,60)[l]{$p_A$} \Text(70,0)[l]{$q$} \Text(115,70)[l]{$k_1$}
\Text(115,30)[l]{$-k_2$} \Text(115,10)[l]{$k$}

\ArrowLine(340,50)(410,30)\ArrowLine(410,70)(340,50)
\Photon(310,50)(337,50){3}{4} \ArrowLine(337,50)(340,50)
\Gluon(365,43)(365,23){3}{2} \ZigZag(365,23)(365,3){3}{2}
\ArrowLine(365,3)(365,0) \Gluon(365,23)(406.5,11){3}{4}
\ArrowLine(406.5,11)(410,10)

\Text(310,60)[l]{$p_A$} \Text(370,0)[l]{$q$}
\Text(415,70)[l]{$-k_2$} \Text(415,30)[l]{$k_1$}
\Text(415,10)[l]{$k$}
\end{picture}
\caption[]{Schematic representation of the vertex
$\Gamma^{c(0)}_{\gamma^*q\bar qg}$.}
\end{figure}
The Reggeon momentum is  given by Eq. (\ref{26}), where $\kappa$
now is the quark-antiquark-gluon squared invariant mass. The
vertex $\Gamma_{\gamma^*q\bar qg}^{c(0)}$ obtained in this way is
invariant with respect  to the gauge transformations of the
emitted gluon polarization vector $e_g$ and can be simplified by
appropriate choice of the gauge. We use the axial gauge
\begin{equation}\label{214}
e_gp_2 = 0,\ \ \ e_g = - \frac{2({e_g}_\perp
k_\perp)}{s\beta}p_2+{e_g}_\perp,
\end{equation}
where  $\beta$ is defined by $k = \beta p_1 + \vec
k^{~2}/(s\beta)p_2 + k_\perp$. In this gauge the last nonlocal
term in the expression for the gluon-Reggeon interaction vertex
of Fig. 3(b) disappears and we obtain
$$
\Gamma_{\gamma^*q\bar qg}^{c(0)}(eq_fg^2)^{-1} = \langle 1|t^bt^c|2
\rangle\left[ \bar u_1\left\{ \frac{1}{(p_A - k_1)^2(k_2 + q)^2}\not
e(\not p_A - \not k_1)\not e_g^*(\not k_2 + \not q)\frac{\not p_2}{s}
\right.\right.
$$
$$
+ \frac{1}{(k + k_1)^2(p_A - k_2)^2}\not e_g^*(\not k + \not k_1)
\frac{\not p_2}{s}(\not p_A - \not k_2)\not e - \frac{1}{(k + k_1)
^2(k_2 + q)^2}\not e_g^*(\not k + \not k_1)\not e\times
$$
$$
\left.\left.\times(\not k_2 + \not q)\frac{\not p_2}{s} + \left(
\frac{\gamma_\mu(\not p_A - \not k_2)\not e}{(p_A - k_2)^2} - \frac
{\not e(\not p_A - \not k_1)\gamma_\mu}{(p_A - k_1)^2} \right)\frac{1}
{(k + q)^2}\left( \beta e_g^{*\mu} - \frac{p_2^\mu}{s}(2qe_g^*) \right)
\right\}v_2 \right]
$$
$$
+ \langle 1|t^ct^b|2\rangle\left[ 1 \leftrightarrow 2 \right] =
$$
$$
= \langle 1|t^bt^c|2\rangle\left[ \bar u_1\left\{ \frac{1}{(p_A - k_1)^2
(k_2 + q)^2}\not e(\not p_A - \not k_1)\not e_g^*(\not k_2 + \not q)
\frac{\not p_2}{s} + \frac{1}{(k + k_1)^2(p_A - k_2)^2}\times\right.\right.
$$
$$
\times\not e_g^*(\not k + \not k_1)\frac{\not p_2}{s}(\not p_A - \not k_2)
\not e - \frac{1}{(k + k_1)^2(k_2 + q)^2}\not e_g^*(\not k + \not k_1)\not
e(\not k_2 + \not q)\frac{\not p_2}{s}
$$
$$
\left.\left. + \left( \frac{\gamma_\mu(\not p_A - \not k_2)\not e}
{(p_A - k_2)^2} - \frac{\not e(\not p_A - \not k_1)\gamma_\mu}{(p_A - k_1)^2}
\right)\frac{1}{(k + q)^2}\left( \beta e_g^{*\mu} - \frac{p_2^\mu}{s}(2qe_g^*)
\right) \right\}v_2 \right]
$$
$$
+ \langle 1|t^ct^b|2\rangle\left[ \bar u_1\left\{ \frac{1}{(p_A - k_2)^2
(k_1 + q)^2}\frac{\not p_2}{s}(\not k_1 + \not q)\not e_g^*(\not p_A -
\not k_2)\not e + \frac{1}{(k + k_2)^2(p_A - k_1)^2}\times\right.\right.
$$
$$
\times\not e(\not p_A - \not k_1)\frac{\not p_2}{s}(\not k + \not k_2)
\not e_g^* - \frac{1}{(k + k_2)^2(k_1 + q)^2}\frac{\not p_2}{s}(\not k_1
+ \not q)\not e(\not k + \not k_2)\not e_g^*
$$
\begin{equation}\label{215}
\left.\left. + \left( \frac{\not e(\not p_A - \not k_1)\gamma_\mu}
{(p_A - k_1)^2} - \frac{\gamma_\mu(\not p_A - \not k_2)\not e}{(p_A
- k_2)^2} \right)\frac{1}{(k + q)^2}\left( \beta e_g^{*\mu} - \frac
{p_2^\mu}{s}(2qe_g^*) \right) \right\}v_2 \right].
\end{equation}

\section{The one-loop correction: the two-gluon exchange diagrams}
\setcounter{equation}{0}

In this section we consider the  contribution of the two gluon
exchange diagrams to $\Gamma_{\gamma^*q\bar q}^{c}$. There are
six diagrams of such kind for the process we consider; they are
shown at Fig. 5.

\begin{figure}[tb]
\begin{center}
\begin{picture}(400,80)(0,-40)

\ArrowLine(0,0)(50,0) \ArrowLine(50,0)(100,0)
\ArrowLine(30,50)(100,70) \ArrowLine(100,30)(30,50)
\Photon(0,50)(27,50){3}{4} \ArrowLine(27,50)(30,50)
\Gluon(40,0)(80,64.29){2}{12} \Gluon(80,0)(50,55.71){2}{10}

\Text(0,60)[l]{$p_A$} \Text(0,-10)[l]{$p_B$}
\Text(100,-10)[r]{$p_{B^\prime}$} \Text(105,70)[l]{$k_1$}
\Text(105,30)[l]{$-k_2$}

\ArrowLine(150,0)(200,0) \ArrowLine(200,0)(250,0)
\ArrowLine(180,50)(250,70) \ArrowLine(250,30)(180,50)
\Photon(150,50)(177,50){3}{4} \ArrowLine(177,50)(180,50)
\Gluon(200,0)(200,55.71){2}{9} \Gluon(230,0)(230,35.71){2}{6}

\Text(150,60)[l]{$p_A$} \Text(150,-10)[l]{$p_B$}
\Text(250,-10)[r]{$p_{B^\prime}$} \Text(255,70)[l]{$k_1$}
\Text(255,30)[l]{$-k_2$}

\ArrowLine(300,0)(350,0) \ArrowLine(350,0)(400,0)
\ArrowLine(330,50)(400,70) \ArrowLine(400,30)(330,50)
\Photon(300,50)(327,50){3}{4} \ArrowLine(327,50)(330,50)
\Gluon(350,0)(380,35.71){2}{6} \Gluon(380,0)(350,44.29){2}{8}

\Text(300,60)[l]{$p_A$} \Text(300,-10)[l]{$p_B$}
\Text(400,-10)[r]{$p_{B^\prime}$} \Text(405,70)[l]{$k_1$}
\Text(405,30)[l]{$-k_2$}

\Text(50,-20)[c]{$(1)$} \Text(200,-20)[c]{$(2)$}
\Text(350,-20)[c]{$(3)$}

\end{picture}
\begin{picture}(400,80)(0,-20)

\ArrowLine(0,0)(50,0) \ArrowLine(50,0)(100,0)
\ArrowLine(30,50)(100,70) \ArrowLine(100,30)(30,50)
\Photon(0,50)(27,50){3}{4} \ArrowLine(27,50)(30,50)
\Gluon(50,0)(80,35.71){2}{6} \Gluon(80,0)(50,55.71){2}{10}

\Text(0,60)[l]{$p_A$} \Text(0,-10)[l]{$p_B$}
\Text(100,-10)[r]{$p_{B^\prime}$} \Text(105,70)[l]{$k_1$}
\Text(105,30)[l]{$-k_2$}

\ArrowLine(150,0)(200,0) \ArrowLine(200,0)(250,0)
\ArrowLine(180,50)(250,70) \ArrowLine(250,30)(180,50)
\Photon(150,50)(177,50){3}{4} \ArrowLine(177,50)(180,50)
\Gluon(200,0)(200,55.71){2}{9} \Gluon(230,0)(230,64.29){2}{10}

\Text(150,60)[l]{$p_A$} \Text(150,-10)[l]{$p_B$}
\Text(250,-10)[r]{$p_{B^\prime}$} \Text(255,70)[l]{$k_1$}
\Text(255,30)[l]{$-k_2$}

\ArrowLine(300,0)(350,0) \ArrowLine(350,0)(400,0)
\ArrowLine(330,50)(400,70) \ArrowLine(400,30)(330,50)
\Photon(300,50)(327,50){3}{4} \ArrowLine(327,50)(330,50)
\Gluon(350,0)(350,44.29){2}{7} \Gluon(380,0)(380,35.71){2}{6}

\Text(300,60)[l]{$p_A$} \Text(300,-10)[l]{$p_B$}
\Text(400,-10)[r]{$p_{B^\prime}$} \Text(405,70)[l]{$k_1$}
\Text(405,30)[l]{$-k_2$}

\Text(50,-20)[c]{$(4)$} \Text(200,-20)[c]{$(5)$}
\Text(350,-20)[c]{$(6)$}

\end{picture}
\caption[]{The two-gluon exchange Feynman diagrams for the process
$\gamma^*Q \rightarrow (q\bar q)Q$.}
\end{center}
\end{figure}
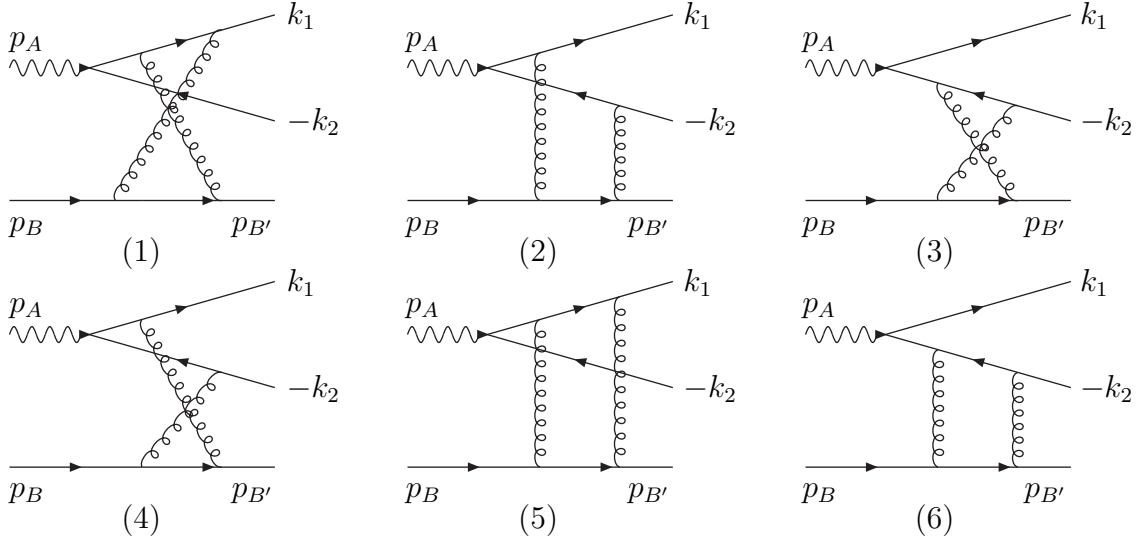
Now we have to  perform the projection on the negative signature
and the octet colour state in the $t$-channel. It is done by the
following replacement of the colour factor of the lowest line of
the diagrams Fig. 5:
\begin{equation}\label{31}
\left( t^bt^a \right)_{B^\prime B} \rightarrow \frac{1}{2}\left(
t^bt^a - t^at^b \right)_{B^\prime B} =
\frac{1}{2}T^c_{ab}t^c_{B^\prime B}~.
\end{equation}
Then we obtain
\begin{equation}\label{32}
{\cal A}_{Q\gamma^* \rightarrow Qq\bar q}^{(2g)(8,-)(1)} = \frac{1}
{4}Nt^c_{i_1i_2}t^c_{B^\prime B}\left\{ \left[ \left( D_1 + D_2 \right)
- \left( 1 \leftrightarrow 2 \right) \right] - \left[ s \leftrightarrow
-s \right] \right\},
\end{equation}
where $N$ is the number of colours, $D_1$ is the amplitude
represented by the diagram of Fig. 5(1) with omitted colour
generators in any its vertex, and $2D_2$ is such amplitude for
the diagram of Fig. 5(2).

The calculation of $D_1$ is quite straightforward. Note here that
since our final goal is the virtual photon impact factor in the
physical space-time, and since the integration over the
quark-antiquark states in Eq. (\ref{13}) is not singular, we need
to retain in  the vertex $\Gamma_{\gamma^*q\bar q}^{c}$ and
consequently in the  amplitude ${\cal A}_{Q\gamma^* \rightarrow
Qq\bar q}^{(2g)(8,-)(1)}$ only the terms which do not vanish at
$\epsilon \rightarrow 0$.  Therefore all one-loop results are
presented in this paper with such accuracy.  In the convenient for
us form we have
$$
\left( D_1 - D_1(1 \leftrightarrow 2) \right) - \left( s \leftrightarrow
-s \right) = \frac{4}{N}g\bar u_{B^\prime}\frac{\not p_1}{s}u_B( - )
eq_fg\bar u_1\left( \frac{\hat \Gamma_1}{t_1} - \frac{\hat \Gamma_2}{t_2}
\right)\frac{\not p_2}{s}v_2\frac{s}{t}\omega^{(1)}(t)
$$
$$
\times\left( \ln\left( \frac{s}{-t} \right) + \ln\left( \frac{-s}{-t}
\right) \right) + 4g\bar u_{B^\prime}\frac{\not p_1}{s}u_B\frac{2s}{t}
eq_fg^3\frac{\Gamma(2-\epsilon)}{(4\pi)^{2+\epsilon}}\frac{1}{2\epsilon}
$$
$$
\times\biggl\{ \biggl[ \bar u_1\frac{\hat \Gamma_1}{t_1}\frac{\not p_2}{s}
v_2\biggl( 2(-t)^\epsilon \left( \frac{1}{\epsilon} + 1 + 2(1+\epsilon)\ln
x_2 + \epsilon - 5\epsilon\psi^\prime(1) \right)
$$
\begin{equation}\label{33}
+ \int_0^1\frac{dy}{\left( - (1-y)t - yt_1 \right)^{1-\epsilon}}
2(1+\epsilon)\left( t - 2(t-t_1)y^\epsilon \right) \biggr)
\biggr] - \biggl[ 1 \leftrightarrow 2 \biggr] \biggr\},
\end{equation}
where the first term is responsible for the Reggeization of the
amplitude ${\cal A}_{Q\gamma^* \rightarrow Qq\bar q}^{(8,-)}$
with $\omega^{(1)}$ being the one-loop Reggeized gluon trajectory,
\begin{equation}\label{o1}
\omega^{(1)}(t)=
-g^2N\frac{\Gamma(1-\epsilon)}{(4\pi)^{2+\epsilon}}\left( \vec q^{~2}
\right)^\epsilon\frac{\Gamma^2(\epsilon)}{\Gamma(2\epsilon)}~,
\end{equation}
and $\Gamma(z)$ and $\psi(z)$ are the Euler $\Gamma$-function and
its logarithmic derivative correspondingly. The calculation of
$D_2$ is more complicated  and we present some details of it in
the Appendix. Here we write down only the result
$$
\left( D_2 - D_2(1 \leftrightarrow 2) \right) - \left( s \leftrightarrow
-s \right) = 4g\bar u_{B^\prime}\frac{\not p_1}{s}u_B\frac{2s}{t}eq_fg^3
\frac{\Gamma(2-\epsilon)}{(4\pi)^{2+\epsilon}}\frac{1}{2\epsilon}
$$
$$
\times\biggl\{ \biggl[ \bar u_1\int_0^1\int_0^1\frac{dy_1dy_2}{\left[
(1-y_2)\left( -(1-y_1)t-y_1t_2 \right) + y_2\left( -(1-y_1)t_1+y_1Q^2
\right) \right]^{2-\epsilon}}\biggl( y_1^{\epsilon-1}(1-y_1)y_2^{-\epsilon}
$$
$$
\times\left( x_1^\epsilon x_2^{-\epsilon} -
2\epsilon^2\psi^\prime(1) \right)2t\hat \Gamma_1 + (1-y_1)4t(ek_1)
$$
\begin{equation}\label{34}
+ \left( y_1^\epsilon y_2^{-\epsilon}x_1^\epsilon x_2^{-\epsilon}
- 1 \right)4x_2t(ep_1) \biggr)\frac{\not p_2}{s}v_2 \biggr] -
\biggl[ 1 \leftrightarrow 2 \biggr] \biggr\}.
\end{equation}
Note that  the imaginary parts of $D_2$  (in the $(p_{B^\prime} +
k_2)^2$-channel) and $D_2(1 \leftrightarrow 2)$  (in the
$(p_{B^\prime} + k_1)^2$-channel) which would destroy the
Reggeization cancel  in the amplitude ${\cal A}_{Q\gamma^*
\rightarrow Qq\bar q}^{(8,-)}$.  Although in the NLA BFKL approach
there is no requirement of the Reggeization of full amplitudes
(the Reggeization of their real parts is sufficient), we see that
nevertheless the Reggeization holds also without omitting of any
imaginary part for  the process  $Q\gamma^* \rightarrow Qq\bar q$.

The amplitude ${\cal A}_{Q\gamma^* \rightarrow Qq\bar q}$ with
the octet colour state and the negative signature in the
$t$-channel has the following Reggeized form
$$
{\cal A}_{Q\gamma^* \rightarrow Qq\bar q}^{(8,-)} = \Gamma_{\gamma^*q\bar q}
^{c}\frac{s}{t}\left[ \left( \frac{s}{-t} \right)^{\omega(t)} + \left(
\frac{-s}{-t} \right)^{\omega(t)} \right]\Gamma_{QQ}^{c} \approx
\Gamma_{\gamma^*q\bar q}^{c(0)}\frac{2s}{t}\Gamma_{QQ}^{c(0)} +
\Gamma_{\gamma^*q\bar q}^{c(0)}\frac{s}{t}\omega^{(1)}(t)
$$
\begin{equation}\label{35}
\times\left[ \ln\left( \frac{s}{-t} \right) + \ln\left(
\frac{-s}{-t} \right) \right]\Gamma_{QQ}^{c(0)} + \Gamma
_{\gamma^*q\bar q}^{c(0)}\frac{2s}{t}\Gamma_{QQ}^{c(1)} +
\Gamma_{\gamma^*q\bar q}^{c(1)}\frac{2s}{t} \Gamma_{QQ}^{c(0)}.
\end{equation}
 Let us now split the one-loop contributions to this
amplitude and both of the effective vertices according to the
three sets of the one-loop diagrams for  ${\cal A}_{Q\gamma^*
\rightarrow Qq\bar q}$: the two-gluon exchange diagrams, the
$t$-channel gluon self-energy diagrams and the one-gluon exchange
diagrams
$$
{\cal A}_{Q\gamma^* \rightarrow Qq\bar q}^{(2g)(8,-)(1)} + {\cal A}
_{Q\gamma^* \rightarrow Qq\bar q}^{(se)(8,-)(1)} + {\cal A}_{Q\gamma
^* \rightarrow Qq\bar q}^{(1g)(8,-)(1)} = \left\{ \Gamma_{\gamma^*q\bar q}
^{(2g)c(1)}\frac{2s}{t}\Gamma_{QQ}^{c(0)} + \Gamma_{\gamma^*q\bar q}^{c(0)}
\frac{2s}{t}\Gamma_{QQ}^{(2g)c(1)} \right.
$$
$$
\left. + \Gamma_{\gamma^*q\bar q}^{c(0)}\frac{s}{t}\omega^{(1)}(t)
\left[ \ln\left( \frac{s}{-t} \right) + \ln\left( \frac{-s}{-t}
\right) \right]\Gamma_{QQ}^{c(0)} \right\}
$$
\begin{equation}\label{36}
+ \left\{ \Gamma_{\gamma^*q\bar q}^{(se)c(1)}\frac{2s}{t}\Gamma
_{QQ}^{c(0)} +  \Gamma_{\gamma^*q\bar q}^{c(0)}\frac{2s}{t}
\Gamma_{QQ}^{(se)c(1)} \right\} +  \left\{ \Gamma_{\gamma^*q\bar q}
^{(1g)c(1)}\frac{2s}{t}\Gamma_{QQ}^{c(0)} + \Gamma_{\gamma^*q\bar q}
^{c(0)}\frac{2s}{t}\Gamma_{QQ}^{(1g)c(1)} \right\},
\end{equation}
where the self-energy diagrams and one-gluon exchange diagrams
have automatically only the octet colour state and negative
signature in the $t$-channel, so that
\begin{equation}\label{37}
{\cal A}_{Q\gamma^* \rightarrow Qq\bar q}^{(se)(8,-)(1)} \equiv
{\cal A}_{Q\gamma^* \rightarrow Qq\bar q}^{(se)(1)},\ \ \
{\cal A}_{Q\gamma^* \rightarrow Qq\bar q}^{(1g)(8,-)(1)} \equiv
{\cal A}_{Q\gamma^* \rightarrow Qq\bar q}^{(1g)(1)}.
\end{equation}
We remind that in our case of completely massless quantum field
theory the contribution from the renormalization of the external
lines is absent in the dimensional regularization.  Now, from the
representations of Eqs. (\ref{36}),  (\ref{37}) it is easy to see
that  $\Gamma_{\gamma^*q\bar q}^{(1g)c(1)}$ is given by the
radiative corrections to the amplitude of the quark-antiquark
production in collision of the virtual photon with the gluon
having momentum $q$, colour index $c$ and polarization vector
$-p_2^\mu/s$, whereas $\Gamma_{QQ}^{(1g)c(1)}$ is defined by the
radiative corrections to the vertex of interaction of this gluon
with the quark $Q$. In both cases the gluon self-energy is not
included into these corrections; it is divided in equal parts
between  $\Gamma_{\gamma^*q\bar q}^{(se)c(1)}$ and
$\Gamma_{QQ}^{(se)c(1)}$ . For the two-gluon exchange
contributions we have the relation:
\begin{equation}\label{38}
\Gamma_{\gamma^*q\bar q}^{(2g)c(1)}\frac{2s}{t}\Gamma_{QQ}^{c(0)} +
\Gamma_{\gamma^*q\bar q}^{c(0)}\frac{2s}{t}\Gamma_{QQ}^{(2g)c(1)} =
{\cal A}_{Q\gamma^* \rightarrow Qq\bar q}^{(2g)(8,-)(1)} - \Gamma
_{\gamma^*q\bar q}^{c(0)}\frac{s}{t}\omega^{(1)}(t)\left[ \ln\left(
\frac{s}{-t} \right) + \ln\left( \frac{-s}{-t}
\right) \right]\Gamma_{QQ}^{c(0)},
\end{equation}
which shows that we need to know the correction
$\Gamma_{QQ}^{(2g)c(1)}$. This correction   can be obtained from
the two-gluon contribution to the $Qq$ elastic scattering
amplitude with the colour octet and the negative signature in the
$t$-channel in the Regge kinematical region. From  Eq. (\ref{38})
with the replacements
\begin{equation}\label{39}
\Gamma_{\gamma^*q\bar q}^{(2g)c(0, 1)} \rightarrow
\Gamma_{qq}^{(2g)c(0, 1)},\ \ \ {\cal A}_{Q\gamma^* \rightarrow
Qq\bar q}^{(2g)(8,-)(1)} \rightarrow {\cal A}_{Qq \rightarrow
Qq}^{(2g)(8,-)(1)},
\end{equation}
denoting
\begin{equation}\label{311} \Gamma_{qq}^{(2g)c(1)} =\delta^{(2g)}(t)
\Gamma_{qq}^{c(0)}, \,\,\, \Gamma_{QQ}^{(2g)c(1)} =
\delta^{(2g)}(t) \Gamma_{QQ}^{c(0)}
\end{equation}
 we get
\begin{equation}\label{310}
\delta^{(2g)}(t) =
\frac{t}{4s\Gamma_{qq}^{c(0)}\Gamma_{QQ}^{c(0)}}\left\{ {\cal
A}_{Qq \rightarrow Qq} ^{(2g)(8,-)(1)} -
\Gamma_{qq}^{c(0)}\Gamma_{QQ}^{c(0)}\frac{s}{t}\omega^{(1)}(t)\left[
\ln\left( \frac{s}{-t} \right) + \ln\left( \frac{-s}{-t} \right)
\right] \right\}.
\end{equation}
The value ${\cal A}_{Qq \rightarrow Qq}^{(2g)(8,-)(1)}$ is given
by the contribution of two diagrams of Fig. 6
\begin{figure}[tb]
\begin{center}
\begin{picture}(250,70)(0,-20)

\ArrowLine(0,0)(50,0) \ArrowLine(50,0)(100,0)
\ArrowLine(0,50)(50,50) \ArrowLine(50,50)(100,50)
\Gluon(30,0)(70,50){2}{8} \Gluon(70,0)(30,50){2}{8}

\Text(0,60)[l]{$p_1$} \Text(100,60)[r]{$p_1^{~\prime}$}
\Text(0,-10)[l]{$p_B$} \Text(100,-10)[r]{$p_{B^\prime}$}

\ArrowLine(150,0)(200,0) \ArrowLine(200,0)(250,0)
\ArrowLine(150,50)(200,50) \ArrowLine(200,50)(250,50)
\Gluon(180,0)(180,50){2}{6} \Gluon(220,0)(220,50){2}{6}

\Text(150,60)[l]{$p_1$} \Text(250,60)[r]{$p_1^{~\prime}$}
\Text(150,-10)[l]{$p_B$} \Text(250,-10)[r]{$p_{B^\prime}$}

\Text(50,-20)[c]{$(a)$} \Text(200,-20)[c]{$(b)$}

\end{picture}
\end{center}
\caption[]{The two-gluon exchange Feynman diagrams for the process
$qQ \rightarrow qQ$.}
\end{figure}
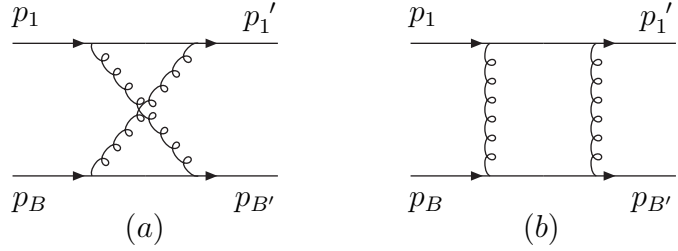
through the same procedure (\ref{31}) for the lowest line of these
diagrams to project on the negative signature and the colour octet
in the $t$-channel. In this way one gets
\begin{equation}\label{312}
{\cal A}_{Qq \rightarrow Qq}^{(2g)(8,-)(1)} = \frac{1}{4}N\langle1
^\prime|t^c|1\rangle\langle B^\prime|t^c|B\rangle
\left( D_3 - D_3(s \leftrightarrow -s) \right),
\end{equation}
where the $D_3$ is the amplitude represented by the diagram of
Fig. 6(a) with omitted colour factors. The calculation of this
diagram is quite  simple and we do not present any details here.
The result for $\delta^{(2g)}$  can be obtained in exact form
without $\epsilon$-expansion
$$
\delta^{(2g)}(t) = \frac{1}{2}\omega^{(1)}(t)\left[ \frac{1}{\epsilon}
+ \psi(1) + \psi(1-\epsilon) - 2\psi(1+\epsilon) \right]
$$
\begin{equation}\label{313}
\approx -g^2N\frac{\Gamma(2-\epsilon)}{(4\pi)^{2+\epsilon}}\frac{1}
{\epsilon}\left( -t \right)^\epsilon\left( \frac{1}{\epsilon} + 1 +
\epsilon - 4\epsilon\psi^\prime(1) \right),
\end{equation}
where the last approximate equality shows the expanded in
$\epsilon$ result which is enough for our purposes. Now, using
Eqs. (\ref{25}), (\ref{27}), (\ref{32}) - (\ref{34}), (\ref{38}),
(\ref{311}) and  (\ref{313}) we obtain
$$
\Gamma_{\gamma^*q\bar q}^{(2g)c(1)}\left(
eq_fg^3Nt^c_{i_1i_2}\frac{\Gamma(2-\epsilon)}{(4\pi)^{2+\epsilon}}
\frac{1}{2\epsilon} \right)^{-1} =  \biggl[ \bar u_1\biggl(
2(-t)^\epsilon\left( 2(1+\epsilon)\ln x_2 - \epsilon\psi^\prime
(1) \right)\frac{\hat \Gamma_1}{t_1}
$$
$$
+\int_0^1\frac{dy}{\left( - (1-y)t - yt_1
\right)^{1-\epsilon}}2(1+\epsilon)\left( t - 2y^\epsilon (
t-t_1)\right)\frac{\hat \Gamma_1}{t_1}
$$
$$
+ \int_0^1\int_0^1\frac{dy_1dy_2}{\left(
-y_1y_2\kappa-t-y_2(t_1-t)-y_1(t_2-t) \right)
^{2-\epsilon}}\biggl\{ y_1^{\epsilon-1}(1-y_1)y_2^{-\epsilon}
$$
$$
\times\left( x_1^\epsilon x_2^{-\epsilon} -
2\epsilon^2\psi^\prime(1) \right)2t\hat \Gamma_1 + (1-y_1)4t(ek_1)
$$
\begin{equation}\label{314}
+ \left( y_1^\epsilon y_2^{-\epsilon}x_1^\epsilon x_2^{-\epsilon}
- 1 \right)4x_2t(ep_1) \biggr\} \biggr)\frac{\not p_2} {s}v_2
\biggr] - \biggl[ 1 \leftrightarrow 2 \biggr].
\end{equation}
The last equality gives the integral representation for the
one-loop two-gluon exchange part of the Reggeon-virtual photon
effective vertex for the quark-antiquark production. Although all
the integrals in (\ref{314}) can be expressed  in terms of
elementary functions and dilogarithms with the necessary accuracy
in $\epsilon$-expansion, it seems  more convenient to leave the
result in such unintegrated form in order to have a possibility to
use usual Feynman parametrization and to change orders of
integrations over all Feynman parameters at  subsequent
calculation of the impact factor. Performing the integrations in
Eq. (\ref{314}) one looses this possibility and has to do a step
back to an unintegrated result to restore it. Let us finally
note, that the method of extraction of  $\Gamma_{\gamma^*q\bar
q}^{(2g)c(1)}$ from the corresponding part of the amplitude we
used here is absolutely equivalent to  one proposed in Ref.
\cite{FM99} and gives the same result that has been checked by
direct comparison.

\section{The one-loop correction: the one-gluon exchange diagrams}
\setcounter{equation}{0}

In this section we consider the contribution of the  one-gluon
exchange diagrams to the vertex $\Gamma_{\gamma^*q\bar q}^c$. It
is presented by the diagrams of Fig. 7
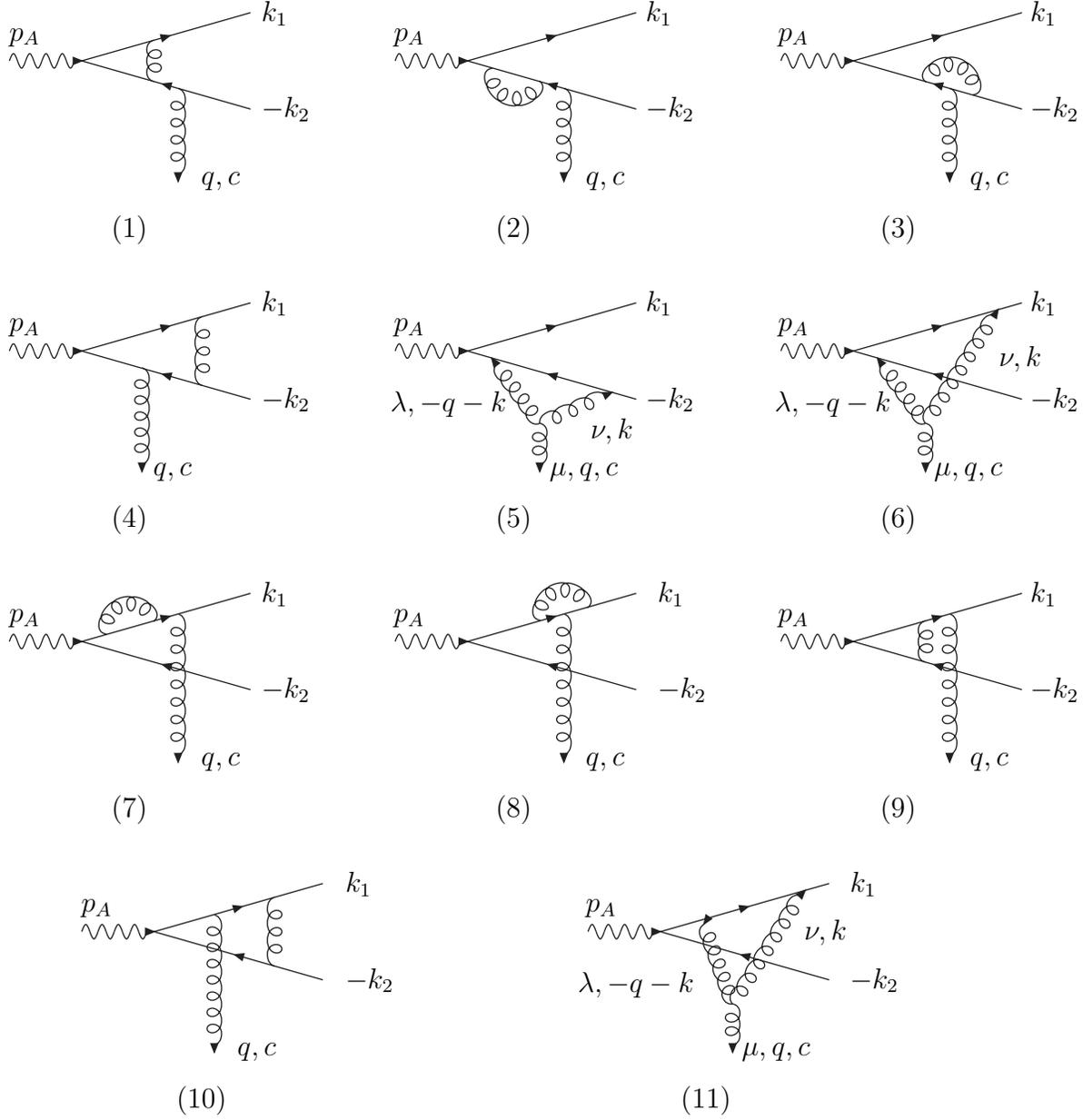
\begin{figure}[tb]
\begin{center}
\begin{picture}(460,100)(10,-75)
\ArrowLine(40,50)(110,70) \ArrowLine(110,30)(40,50)
\Photon(10,50)(37,50){3}{4} \ArrowLine(37,50)(40,50)
\Gluon(80,38.57)(80,3){3}{4} \ArrowLine(80,3)(80,0)
\Gluon(70,41.43)(70,58.57){3}{2}

\Text(10,60)[l]{$p_A$} \Text(90,0)[l]{$q, c$}
\Text(115,70)[l]{$k_1$} \Text(115,30)[l]{$-k_2$}
\ArrowLine(200,50)(270,70) \ArrowLine(270,30)(200,50)
\Photon(170,50)(197,50){3}{4} \ArrowLine(197,50)(200,50)
\Gluon(240,38.57)(240,3){3}{4} \ArrowLine(240,3)(240,0)
\GlueArc(220,44.28)(10,165,345){3}{4}

\Text(170,60)[l]{$p_A$} \Text(250,0)[l]{$q, c$}
\Text(275,70)[l]{$k_1$} \Text(275,30)[l]{$-k_2$}
\ArrowLine(360,50)(430,70) \ArrowLine(430,30)(360,50)
\Photon(330,50)(357,50){3}{4} \ArrowLine(357,50)(360,50)
\Gluon(400,38.57)(400,3){3}{4} \ArrowLine(400,3)(400,0)
\GlueArc(400,38.57)(10,-15,165){3}{4}

\Text(330,60)[l]{$p_A$} \Text(410,0)[l]{$q, c$}
\Text(435,70)[l]{$k_1$} \Text(435,30)[l]{$-k_2$}

\Text(60,-20)[c]{$(1)$} \Text(220,-20)[c]{$(2)$}
\Text(380,-20)[c]{$(3)$}

\end{picture}
\begin{picture}(460,100)(10,-55)
\ArrowLine(40,50)(110,70) \ArrowLine(110,30)(40,50)
\Photon(10,50)(37,50){3}{4} \ArrowLine(37,50)(40,50)
\Gluon(65,43)(65,3){3}{5} \ArrowLine(65,3)(65,0)
\Gluon(90,35.71)(90,64.29){3}{3}

\Text(10,60)[l]{$p_A$} \Text(70,0)[l]{$q, c$}
\Text(115,70)[l]{$k_1$} \Text(115,30)[l]{$-k_2$}

\ArrowLine(200,50)(270,70) \ArrowLine(270,30)(200,50)
\Photon(170,50)(197,50){3}{4} \ArrowLine(197,50)(200,50)
\Gluon(230,20)(230,3){3}{2} \ArrowLine(230,3)(230,0)
\Gluon(230,20)(212,44.44){3}{4} \ArrowLine(212,44.44)(210,47.14)
\Gluon(230,20)(257,31.56){3}{3} \ArrowLine(257,31.56)(260,32.86)

\Text(170,60)[l]{$p_A$} \Text(275,70)[l]{$k_1$}
\Text(275,30)[l]{$-k_2$} \Text(235,0)[l]{$\mu, q, c$}
\Text(252,17)[l]{$\nu, k$} \Text(217,28)[r]{$\lambda,-q-k$}

\ArrowLine(360,50)(430,70) \ArrowLine(430,30)(360,50)
\Photon(330,50)(357,50){3}{4} \ArrowLine(357,50)(360,50)
\Gluon(390,20)(390,3){3}{2} \ArrowLine(390,3)(390,0)
\Gluon(390,20)(372,44.44){3}{4} \ArrowLine(372,44.44)(370,47.14)
\Gluon(390,20)(418.50,64.79){3}{7}
\ArrowLine(418.50,64.79)(420,67.14)

\Text(330,60)[l]{$p_A$} \Text(435,70)[l]{$k_1$}
\Text(435,30)[l]{$-k_2$} \Text(395,0)[l]{$\mu,q, c$}
\Text(422,47)[l]{$\nu, k$} \Text(377,28)[r]{$\lambda, -q-k$}

\Text(60,-20)[c]{$(4)$} \Text(220,-20)[c]{$(5)$}
\Text(380,-20)[c]{$(6)$}

\end{picture}
\begin{picture}(460,100)(10,-35)
\ArrowLine(40,50)(110,70) \ArrowLine(110,30)(40,50)
\Photon(10,50)(37,50){3}{4} \ArrowLine(37,50)(40,50)
\Gluon(80,61.43)(80,3){3}{7} \ArrowLine(80,3)(80,0)
\GlueArc(60,55.72)(10,15,195){3}{4}

\Text(10,60)[l]{$p_A$} \Text(115,70)[l]{$k_1$}
\Text(115,30)[l]{$-k_2$} \Text(90,0)[l]{$q, c$}

\ArrowLine(200,50)(270,70) \ArrowLine(270,30)(200,50)
\Photon(170,50)(197,50){3}{4} \ArrowLine(197,50)(200,50)
\Gluon(240,61.43)(240,3){3}{7} \ArrowLine(240,3)(240,0)
\GlueArc(240,61.43)(10,15,195){3}{4}

\Text(170,60)[l]{$p_A$} \Text(280,70)[l]{$k_1$}
\Text(280,30)[l]{$-k_2$} \Text(250,0)[l]{$q, c$}

\ArrowLine(360,50)(430,70) \ArrowLine(430,30)(360,50)
\Photon(330,50)(357,50){3}{4} \ArrowLine(357,50)(360,50)
\Gluon(400,61.43)(400,3){3}{7} \ArrowLine(400,3)(400,0)
\Gluon(390,41.43)(390,58.57){3}{2}

\Text(330,60)[l]{$p_A$} \Text(435,70)[l]{$k_1$}
\Text(435,30)[l]{$-k_2$} \Text(410,0)[l]{$q, c$}

\Text(60,-20)[c]{$(7)$} \Text(220,-20)[c]{$(8)$}
\Text(380,-20)[c]{$(9)$}

\end{picture}
\begin{picture}(460,100)(10, -15)
\ArrowLine(70,50)(140,70) \ArrowLine(140,30)(70,50)
\Photon(40,50)(67,50){3}{4} \ArrowLine(67,50)(70,50)
\Gluon(95,57)(95,3){3}{7} \ArrowLine(95,3)(95,0)
\Gluon(120,35.71)(120,64.29){3}{3}

\Text(40,60)[l]{$p_A$} \Text(150,70)[l]{$k_1$}
\Text(150,30)[l]{$-k_2$} \Text(105,0)[l]{$q, c$}
\ArrowLine(280,50)(350,70) \ArrowLine(350,30)(280,50)
\Photon(250,50)(277,50){3}{4} \ArrowLine(277,50)(280,50)
\Gluon(310,20)(310,3){3}{2} \ArrowLine(310,3)(310,0)
\Gluon(310,20)(298,53){3}{5} \ArrowLine(300,56)(298,53)
\Gluon(310,20)(338,65){3}{7} \ArrowLine(338,65)(340,67.14)

\Text(250,60)[l]{$p_A$} \Text(360,70)[l]{$k_1$}
\Text(360,30)[l]{$-k_2$} \Text(315,0)[l]{$\mu,q, c$}
\Text(341,50)[l]{$\nu, k$} \Text(295,28)[r]{$\lambda, -q-k$}

\Text(90, -20)[c]{$(10)$} \Text(300, -20)[c]{$(11)$}

\end{picture}
\end{center}
\caption[]{The diagrams corresponding to the correction
$\Gamma_{\gamma^*q\bar q}^{(1g)c(1)}$.}
\end{figure}
with the gluon polarization vector equal to $-p_2^\mu/s$,  as
already was explained in the previous Section. Calculating the
colour factors of the diagrams one can easily obtain the
following representation
\begin{equation}\label{41}
\Gamma_{\gamma^*q\bar q}^{(1g)c(1)} = Nt^c_{i_1i_2}\left\{ \biggl[
-\frac{2C_F}{N}\left( R_1 + R_2 \right) + \frac{N - 2C_F}{N}\left(
R_3 + R_4 \right) + R_5 + \tilde R_6 \biggr] - \biggl[
1 \leftrightarrow 2 \biggr] \right\},
\end{equation}
with the usual notation
\begin{equation}\label{42}
C_F= \frac{N^2 - 1}{2N}
\end{equation}
and the notations $2R_1, ... , 2R_4, -2R_5$ and $4\tilde R_6$ for
the amplitudes represented by the diagrams of Figs. 7(1), ... ,
(4), (5) and (6) respectively with  omitted colour generators in
any vertex and the external virtual gluon polarization vector
equal to $p_2^\mu/s$. While the definition of $R_1, ... , R_4$ is
absolutely clear, the $R_5$ and $\tilde R_6$ are not well defined
by the above prescription because of presence of three-gluon
vertices in the corresponding diagrams Figs. 7(5) and 7(6). To
complete their definition we have indicated explicitly the momenta
and vector indices for the three-gluon vertex for which the
following expression should be used after the omission of colour
generators
\begin{equation}\label{43}
\gamma_{\lambda\nu\mu}(-k-q, k)=ig\left[
-g_{\lambda\nu}(2k+q)_\mu + g_{\lambda\mu}(k+2q)_\nu +
g_{\nu\mu}(k-q)_\lambda \right].
\end{equation}

The calculation of $R_1, R_2, R_3$ and $R_5$ is simple and we
present here only the list of the results in integral form
without any details
$$
R_1 = -
eq_fg^3\frac{\Gamma(2-\epsilon)}{(4\pi)^{2+\epsilon}}\frac{1}{2\epsilon}
\bar u_1\int_0^1\frac{dy} {\left( (1-y)Q^2 - yt_1
\right)^{1-\epsilon}}\biggl\{ (1+2\epsilon)\frac{Q^2}{t_1}\hat
\Gamma_1 + 2\epsilon (ep_1)
$$
\begin{equation}\label{44}
+ y\biggl[ (1-2\epsilon)\left( \frac{Q^2}{t_1} + 1 \right)\hat
\Gamma_1 + 2(2-\epsilon)(ek_1)  \biggr] \biggr\}\frac{\not
p_2}{s}v_2,
\end{equation}
\begin{equation}\label{45}
R_2 = - eq_fg^3\frac{\Gamma(2-\epsilon)}{(4\pi)^{2+\epsilon}}\frac{1}
{2\epsilon}\bar u_1\frac{\hat \Gamma_1}{\left( -t_1 \right)^{1-\epsilon}}
\frac{\not p_2}{s}v_2,
\end{equation}
$$
R_3 = eq_fg^3\frac{\Gamma(2-\epsilon)}{(4\pi)^{2+\epsilon}}\frac{1}
{2\epsilon}\bar u_1\int_0^1\frac{dy}{\left( - (1-y)t - yt_1 \right)
^{1-\epsilon}}\biggl\{ (1+2\epsilon)\frac{t}{t_1}\hat \Gamma_1
$$
\begin{equation}\label{46}
+ y\biggl[ (1-2\epsilon)\left( \frac{t}{t_1} - 1 \right)\hat
\Gamma_1 + (2-\epsilon)x_2\left( \hat \Gamma_1 + \hat \Gamma_2 -
2(e_\perp q_\perp)\right) \biggr] \biggr\}\frac{\not p_2}{s}v_2,
\end{equation}
$$
R_5 =
eq_fg^3\frac{\Gamma(2-\epsilon)}{(4\pi)^{2+\epsilon}}\frac{1}{2\epsilon}\bar
u_1\int_0^1\frac{dy} {\left( - (1-y)t - yt_1
\right)^{1-\epsilon}}\biggl\{ -(1-2\epsilon)\frac{t}{t_1}\hat
\Gamma_1 - (2+\epsilon)\hat \Gamma_1 + (1+2\epsilon)x_2
$$
\begin{equation}\label{47}
\times\left( \hat \Gamma_1 + \hat \Gamma_2 - 2(e_\perp q_\perp)
\right) + y\biggl[ \left( \frac{t}{t_1} - 1 \right)\hat \Gamma_1 -
(1+\epsilon)x_2\left( \hat \Gamma_1 + \hat\Gamma_2 - 2(e_\perp
q_\perp)
 \right) \biggr] \biggr\}\frac{\not p_2}{s}v_2.
\end{equation}

The calculation of $R_4$ and $\tilde R_6$ is, actually, also
quite straightforward, though it is very tedious. To simplify
representation of the results, we split the corresponding
diagrams into two parts: infrared divergent part and convergent
one. For the first part we present results in the spinor
representation and for another part in the helicity
representation with the use of the definitions of Eqs.
(\ref{210}), (\ref{211}). Besides that, instead of the result for
$\tilde R_6$ we present the result for $R_6$
\begin{equation}\label{48}
\tilde R_6 = \frac{1}{2}\left( R_6 - R_6(1 \leftrightarrow 2) \right)
\end{equation}
which can evidently be used in (\ref{41}) instead of $\tilde R_6$.
The results for the singular parts are
$$
R_4^{(s)} = eq_fg^3\frac{\Gamma(2-\epsilon)}{(4\pi)^{2+\epsilon}}
\bar u_1\times
$$
$$
\int_0^1\int_0^1\int_0^1\frac{dzdy_1dy_2\theta\left( 1-y_1-y_2
\right)z^{1+\epsilon}\kappa}{\left[ zy_1y_2\left( -\kappa - i\delta
\right) + (1-z)\left( y_1Q^2 - y_2t - (1-y_1-y_2)t_1 \right) \right]
^{2-\epsilon}}\left[ (1-y_1-y_2)\times\right.
$$
\begin{equation}\label{49}
\left.\left( \left( 1-z(1-y_2) \right)x_2\left( \hat \Gamma_1 +
\hat \Gamma_2 - 2(e_\perp q_\perp)  \right) - \hat \Gamma_1
\right) - 2y_1(1-zy_1) (ek_1) \right]\frac{\not p_2}{s}v_2,
\end{equation}
$$
R_6^{(s)} =
eq_fg^3\frac{\Gamma(2-\epsilon)}{(4\pi)^{2+\epsilon}}\frac{1}{2\epsilon}
\int_0^1\int_0^1\frac{dy_1dy_2(1-y_1)}{\left(
-y_1y_2\kappa-t-y_2(t_1-t)-y_1(t_2-t)\right) ^{2-\epsilon}}
$$
\begin{equation}\label{410}
\times\bar u_1\biggl\{ 2(t_2-t)\left(\hat\Gamma_1
+2y_1(ek_1)\right)- x_1((1-y_1)t_1-y_1Q^2)\left( \hat \Gamma_1 +
\hat \Gamma_2 - 2(e_\perp q_\perp )  \right) \biggr\}\frac{\not
p_2}{s}v_2.
\end{equation}
The results for the regular parts are
$$
R_4^{(r)} = \frac{eq_fg^3}{(4\pi)^2}\frac{2}{\sqrt{2x_1x_2\vec
q^{~2}}}\times
$$
$$
\int_0^1\int_0^1\int_0^1\frac{dzdy_1dy_2\theta\left( 1-y_1-y_2
\right)z}{\left[ zy_1y_2\left( -\kappa - i\delta \right) + (1-z)
\left( y_1Q^2 - y_2t - (1-y_1-y_2)t_1 \right) \right]^2}
\biggl\{ \biggl[ (1-z)\left( y_1Q^2 -\right.
$$
$$
\left.y_2t - (1-y_1-y_2)t_1 \right) - zy_1y_2\kappa
\biggr]\biggl[ \left( (1-y_2)(\vec k_1\vec q+i\lambda P) -
y_2x_1\vec q^{~2} \right)zx_2\delta_{\lambda, -\xi} - \left(
(1-zy_1)\times\right.
$$
$$
\left.\left( x_1\vec q^{~2} + \vec k_1\vec q+i\lambda P \right)
+  (1-z)x_2\vec q^{~2} \right) x_1\delta_{\lambda, \xi} \biggr] +
y_1x_2\kappa\biggl[ \left( 1 -z (1-y_2) \right)x_1\left(
\sqrt{2}qQx_1 \delta_{\lambda, 0}\right.
$$
$$
\left. - (\vec k_1\vec q+i\lambda P)(\delta_{\lambda, \xi} +
\delta_{\lambda, -\xi}) \right) - zy_2x_1 \vec
q^{~2}\delta_{\lambda, -\xi} \biggr] + (1-z)\biggl[
(1-y_1-y_2)\left( \left( x_2\vec q^{~4} + \right.\right.
$$
$$
\left.\left.((\vec k_1-\vec k_2)\vec q)(\vec k_2\vec q-i\lambda
P) \right)x_1\delta_{\lambda, \xi} + x_2\vec k_1^{~2}\vec
q^{~2}(\delta_{\lambda, -\xi} - \delta_{\lambda, \xi}) \right) +
y_2x_1 \vec q^{~2}\left( x_2\vec q^{~2}\delta_{\lambda, -\xi}
-\right.
$$
$$
\left.(\vec k_2\vec q-i\lambda P)\delta_{\lambda, \xi} \right)
\biggr] + (1-z)\left( y_1x_1+(1-y_2)x_2 \right)\biggl[ \left(
x_1(\vec k_2\vec q-i\xi P) - zy_1(\vec k_1\vec q-i\xi P - x_1t_1)
\right)
$$
$$
\times\sqrt{2}qQ\delta_{\lambda, 0} + z(1-y_1-y_2)\left( \left(
\vec k_1^{~2}\vec q^{~2} - x_1t_1 (\vec k_1\vec q+i\lambda P)
\right)(\delta_{\lambda, -\xi}  - \delta_{\lambda, \xi}) + 2(\vec
k_1 \vec k_2 - x_1x_2Q^2)\right.
$$
$$
\left.\times(\vec k_1\vec q+i\lambda P)\delta_{\lambda, -\xi} -
(\vec k_1\vec q-i\xi P - x_1t_1) \sqrt{2}qQx_2\delta_{\lambda, 0}
\right)
$$
\begin{equation}\label{411}
+ zy_2\vec q^{~2}\left( (\vec k_1\vec q + i\lambda P -
x_1t_1)(\delta_{\lambda, -\xi} - \delta_{\lambda, \xi}) - 2(\vec
k_1\vec k_2 - x_1x_2Q^2)\delta_{\lambda, \xi} \right) \biggr]
\biggr\}
\end{equation}
and
$$
R_6^{(r)} = \frac{eq_fg^3}{(4\pi)^2\sqrt{2x_1x_2\vec
q^{~2}}}\int_0^1\int_0^1\!\!\!\frac{dy_1dy_2}{\left(
-y_1y_2\kappa-t-y_2(t_1-t)-y_1(t_2-t) \right)^2}
$$
$$
\times\biggl( y_1\delta_{\lambda, 0}\sqrt{2}qQx_1\biggl\{ \left(
x_1x_2Q^2 - \vec k_1\vec k_2 - i\xi P \right)\left( 1-3x_2
\right) + 2x_2(t_2-t) \biggr\} + (1-y_1)
$$
$$
\times\biggl\{ \biggl[ 2\left(
-y_1y_2\kappa-t-y_2(t_1-t)-y_1(t_2-t) \right) + y_1(t_2-t-\kappa)
\biggr]x_1x_2
$$
$$
\times\biggl( (\delta_{\lambda, -\xi} + \delta_{\lambda,
\xi})(\vec k_1\vec q + i\lambda P) - \delta_{\lambda, 0}
\sqrt{2}qQx_1 \biggr) + \biggl[ 2x_1t_1 + 2x_2t_2 - 2t -
y_2x_2(t_2-t-\kappa) \biggr]
$$
$$
\times\biggl( \delta_{\lambda, 0}\sqrt{2}qQx_1x_2 -
\delta_{\lambda, -\xi}x_2(\vec k_1\vec q + i\lambda P) -
\delta_{\lambda, \xi}x_1(\vec k_2\vec q - i\lambda P) \biggr)
$$
$$
- \biggl[ t\left( 2x_1\delta_{\lambda, \xi} +
3x_2\delta_{\lambda, -\xi} \right) + y_1x_1\left(
t\delta_{\lambda, -\xi} - \delta_{\lambda, 0}\sqrt{2}qQx_1 \right)
\biggr]\left( x_1x_2Q^2 - \vec k_1\vec k_2 - i\xi P \right)
$$
$$
+ \biggl[ 3\left( x_2^2t_2\delta_{\lambda, -\xi} -
x_1^2t_1\delta_{\lambda, \xi} \right) + y_1x_1x_2\left( t_2\delta
_{\lambda, -\xi} - t_1\delta_{\lambda, \xi} \right) -
y_1x_1Q^2\delta_{\lambda, \xi} \biggr](\vec k_1\vec q + i\lambda
P)
$$
\begin{equation}\label{412}
+ x_1t\biggl[ 3x_2^2Q^2\delta_{\lambda, -\xi} - 3\vec
k_1^{~2}\delta_{\lambda, \xi} + y_1x_1Q^2\left( \delta _{\lambda,
-\xi}x_2 + \delta_{\lambda, \xi}x_1 \right) +
2y_2x_2\kappa\delta_{\lambda, \xi} \biggr] \biggr\} \biggr).
\end{equation}

The result for the correction $\Gamma_{\gamma^*q\bar
q}^{(1g)c(1)}$ is obtained now by Eq. (\ref{41}) with the
replacement
\begin{equation}\label{413}
\tilde R_6 \rightarrow R_6,
\end{equation}
where $R_1, R_2, R_3$ and $R_5$ are presented by Eqs. (\ref{44}),
(\ref{45}), (\ref{46}) and (\ref{47}) respectively and $R_4$ and
$R_6$ are given by Eqs. (\ref{49}) - (\ref{412}) and by the
 relation
\begin{equation}\label{414}
R_{4, 6} = R_{4, 6}^{(s)} + R_{4, 6}^{(r)}.
\end{equation}

\section{The total one-loop correction}
\setcounter{equation}{0}

There is one more one-loop contribution to the vertex
$\Gamma_{\gamma^*q\bar q}^c$ related to the  $t$-channel gluon
self-energy.  It is given by the half of the amplitude
schematically represented at Fig. 8
\begin{figure}[tb]
\begin{center}
\begin{picture}(150,80)(0,-20)

\ArrowLine(30,50)(100,70) \ArrowLine(100,30)(30,50)
\Photon(0,50)(27,50){3}{4} \ArrowLine(27,50)(30,50)
\GCirc(40,50){10}{0.5} \Gluon(40,40)(40,20){3}{2}
\GCirc(40,10){10}{0.5} \Gluon(40,0)(40,-17){3}{2}
\ArrowLine(40,-17)(40,-20)

\Text(0,60)[l]{$p_A$} \Text(105,70)[l]{$k_1$}
\Text(105,30)[l]{$-k_2$} \Text(45,-20)[l]{$q,\, c$}

\end{picture}
\end{center}
\caption[]{Schematic representation of  the correction
$\Gamma_{\gamma^*q\bar q}^{(se)c(1)}$.}
\end{figure}
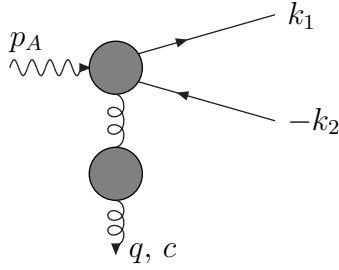
with the gluon polarization vector equal to $-p_2^\mu/s$,  as
was already explained in the Section 3. To find the correction
$\Gamma_{\gamma^*q\bar q}^{(se)c(1)}$  one should only know the
one-loop gluon vacuum polarization  and the Born Reggeon-virtual
photon vertex. We obtain:
$$
\Gamma_{\gamma^*q\bar q}^{(se)c(1)}\left( eq_fg^3Nt^c_{i_1i_2}
\frac{\Gamma(2-\epsilon)}{(4\pi)^{2+\epsilon}}
\frac{1}{2\epsilon} \right)^{-1}
$$
\begin{equation}\label{51}
= \biggl[ \bar u_1\left( -t \right)^\epsilon\biggl\{ \frac{5}{3} -
\frac{2}{3}\frac{n_f}{N} + \epsilon\biggl( \frac{4}{9}\frac{n_f}{N}
- \frac{16}{9} \biggr) \biggr\}\frac{\hat\Gamma_1}{t_1}\frac
{\not p_2}{s}v_2 \biggr] - \biggl[ 1 \leftrightarrow 2 \biggr],
\end{equation}
where  $n_f$ is the number of quark flavours.

We present the total one-loop correction to the vertex of the
quark-antiquark production in the virtual photon-Reggeized gluon
collision in the form:
\begin{equation}\label{52}
\Gamma_{\gamma^*q\bar q}^{c(1)} = \Gamma_{\gamma^*q\bar q}^{(sing)c(1)}
+ \Gamma_{\gamma^*q\bar q}^{(reg)c(1)},
\end{equation}
with
\begin{equation}\label{53}
\Gamma_{\gamma^*q\bar q}^{(reg)c(1)} = Nt^c_{i_1i_2}\biggl\{ \biggl[
\frac{N - 2C_F}{N}R_4^{(r)} + R_6^{(r)} \biggr] - \biggl[
1 \leftrightarrow 2 \biggr] \biggr\}
\end{equation}
where $R_4^{(r)}$ and $R_6^{(r)}$ are given by Eqs. (\ref{411})
and  (\ref{412}) correspondingly,  and
$$
\Gamma_{\gamma^*q\bar q}^{(sing)c(1)} = \Gamma_{\gamma^*q\bar q}
^{(2g)c(1)} + \Gamma_{\gamma^*q\bar q}^{(se)c(1)}
$$
\begin{equation}\label{54}
+ Nt^c_{i_1i_2}\left\{ \biggl[ -\frac{2C_F}{N}\left( R_1 + R_2 \right)
+ \frac{N - 2C_F}{N}\left( R_3 + R_4^{(s)} \right) + R_5 + R_6^{(s)}
\biggr] - \biggl[ 1 \leftrightarrow 2 \biggr] \right\}
\end{equation}
where $\Gamma_{\gamma^*q\bar q}^{(2g)c(1)}, \Gamma_{\gamma^*q\bar
q}^{(se)c(1)}, R_1 - R_3, R_5, R_4^{(s)}$ and $R_6^{(s)}$ are
defined  in  Eqs. (\ref{314}), (\ref{51}),  (\ref{44}) -
(\ref{46}), (\ref{47}), (\ref{49}) and (\ref{410}) respectively.
Using the last set of relations one can easily obtain
$$
\Gamma_{\gamma^*q\bar q}^{(sing)c(1)}\left(
eq_fg^3Nt^c_{i_1i_2}\frac{\Gamma(2-\epsilon)}{(4\pi)^{2+\epsilon}}
\frac{1}{2\epsilon} \right)^{-1} = \biggl[ \bar u_1\biggl( \left(
-t \right)^\epsilon\biggl\{ \frac{5}{3} - \frac{2}{3}
\frac{n_f}{N} + 4(1+\epsilon)\ln x_2 + 2\epsilon\biggl(
\frac{2}{9}\frac{n_f}{N}
$$
$$
- \frac{8}{9} - \psi^\prime(1) \biggr)
\biggr\}\frac{\hat\Gamma_1}{t_1} +
\frac{2C_F}{N}\frac{\hat\Gamma_1} {\left( -t_1
\right)^{1-\epsilon}} + \frac{2C_F}{N}\int_0^1\frac{dy}{\left(
(1-y)Q^2 - yt_1 \right)^{1-\epsilon}} \biggl\{
(1+2\epsilon)\frac{Q^2}{t_1}\hat \Gamma_1 + 2\epsilon (ep_1)
$$
$$
+ y\biggl[ (1-2\epsilon)\left( \frac{Q^2}{t_1} + 1 \right)\hat
\Gamma_1 + 2(2-\epsilon)\left(ek_1\right) \biggr] \biggr\} +
\frac{1}{N}\int_0^1\frac{dy}{\left( -(1-y)t-yt_1
\right)^{1-\epsilon}}
$$
$$
\times\biggl\{ 2\left( (1+3\epsilon)N - (1+2\epsilon)C_F
\right)\frac{t}{t_1}\hat\Gamma_1 - (2+\epsilon)N\hat \Gamma_1 +
(1+2\epsilon)Nx_2\left( \hat\Gamma_1 + \hat\Gamma_2 - 2(e_\perp
q_\perp) \right)
$$
$$
- y^\epsilon4(1+\epsilon)N\left( \frac{t}{t_1} - 1 \right)\hat
\Gamma_1 + y\biggl[ 2\left( (1-\epsilon)N - (1-2\epsilon)
C_F \right)\left( \frac{t}{t_1} - 1 \right)\hat\Gamma_1 +
\left( (1-2\epsilon)N \right.
$$
$$
\left. - 2(2-\epsilon)C_F \right)x_2\left( \hat\Gamma_1 +
\hat\Gamma_2 - 2(e_\perp q_\perp) \right) \biggr] \biggr\} +
\frac{N-2C_F}{N}\times
$$
$$
\int_0^1\int_0^1\int_0^1\frac{dzdy_1dy_2\theta\left( 1-y_1-y_2
\right)z^{1+\epsilon}2\epsilon\kappa}{\left[ zy_1y_2\left(
-\kappa - i\delta \right) + (1-z)\left( y_1Q^2 - y_2t - (1-y_1-y_2)
t_1 \right) \right]^{2-\epsilon}}\left[ (1-y_1-y_2)\times\right.
$$
$$
\left.\left( \left( 1-z(1-y_2) \right)x_2\left( \hat \Gamma_1 +
\hat \Gamma_2 - 2(e_\perp q_\perp) \right) - \hat \Gamma_1 \right)
- 2y_1(1-zy_1)\left(ek_1\right) \right]
$$
$$
+ \int_0^1\int_0^1\frac{dy_1dy_2}{\left( -y_1y_2\kappa-t-y_2(t_1-t)-y_1
(t_2-t) \right)^{2-\epsilon}}\biggl\{ y_1^{\epsilon-1}(1-y_1)y_2^{-\epsilon}
$$
$$
\times\left( x_1^\epsilon x_2^{-\epsilon} -
2\epsilon^2\psi^\prime(1) \right)2t\hat \Gamma_1 + \left(
y_1^\epsilon y_2^{-\epsilon}x_1^\epsilon x_2^{-\epsilon} - 1
\right)4x_2t(ep_1)+ (1-y_1)\biggl[ 2(t_2-t)\hat\Gamma_1
$$
$$
- x_1t_1\left( \hat \Gamma_1 + \hat \Gamma_2 - 2(e_\perp q_\perp)
\right) + 4t(ek_1) \biggr] + y_1(1-y_1)\biggl[ 4(t_2 -t)(ek_1)
$$
\begin{equation}\label{55}
+ x_1(t_1+Q^2)\left( \hat\Gamma_1 + \hat\Gamma_2 - 2(e_\perp
q_\perp) \right) \biggr] \biggr\} \biggr)\frac{\not p_2} {s}v_2
\biggr] - \biggl[ 1 \leftrightarrow 2 \biggr].
\end{equation}

The relations (\ref{52}), (\ref{53}), (\ref{55}) together with
Eqs. (\ref{411}, \ref{412}) present the one-loop correction to the
$\gamma^*R\rightarrow q\bar q$ vertex.

\section{Discussion}
\setcounter{equation}{0}

In this paper we have calculated the effective vertices for the
Reggeon-virtual photon interaction. Starting  from  already known
expressions (\ref{27}), (\ref{28}) for the $q\bar q$ production
vertex in the Born approximation  we have represented this vertex
in the helicity basis (\ref{212}) and then have obtained in the
same approximation  the vertex (\ref{215}) for the $q\bar qg$
production.  The most of the paper is devoted to the calculation
of the one-loop corrections for the $q\bar q$ production vertex,
which are presented in  Eqs. (\ref{52}), (\ref{55}), (\ref{53}),
(\ref{411}) and (\ref{412}). In order to simplify the
presentation the last three results are given in  the helicity
basis, that caused the representation  in this basis also the
Born $q\bar q$ production vertex (\ref{212}).

The obtained results  can be used for theoretical analysis of a
number of processes related to the quark-antiquark production in
the photon fragmentation region. In particular, they are necessary
for  calculation of the virtual photon impact factor at the
next-to-leading order (see Eq. (\ref{13})). We have used the
integral representation for a part of the one-loop corrections to
the  $q\bar q$ production vertex since it is convenient  for
further calculation of the virtual photon impact factor in the
next-to-leading order.

Note, that everywhere in the paper  $g$ is the unrenormalized
coupling constant, so that the one-loop correction contains the
ultraviolet singularities in $\epsilon$. In order to remove them
one should only express $g$ in terms of the renormalized coupling
constant $g(\mu)$. After the renormalization there still remain
the infrared singularities, which must cancel in physical
quantities (in the virtual photon impact factor they cancel
\cite{FM99} with corresponding terms in the contributions from
the additional gluon emission and from the counterterm (see Eq.
(\ref {13})).

Recently an independent calculation of the 
$\gamma^*\to q\bar q$ vertex was reported in the paper
\cite{Bartels:2001gt}. The results of  \cite{Bartels:2001gt}
are presented in the form where all integrations are performed.
At the moment we can only say that these result are very long and 
complicated. It needs definitely some time to make a comparison 
between our and their results.

\vskip 1.5cm \underline {Acknowledgment}: 
This work is partly supported by the
INTAS (00-0036 and 00-00679) 
and by the Russian Fund of Basic Researches
(99-02-16822, 99-02-17211, 00-15-96691 and 01-02-16042).
Two of us (M.K. and D.I.) thank
the Dipartimento di Fisica della Universit\`a della Calabria for
the warm hospitality while part of this work was done.
D.I. was supported by Alexander von Humboldt Stiftung.

\appendix

\section{Appendix A}

In this section we explain very briefly important steps for calculation
of the most complicated diagram $D_2$ (Fig. 5(2)) to make our results
checkable step by step. According to its
definition we obtain from the corresponding diagram
$$
D_2 = \frac{-ieq_fg^4}{2(2\pi)^D}\times
$$
\begin{equation}\label{a1}
\int\frac{d^Dk\bar u_1\gamma^\mu\left( \not k + \not p_A - \not k_2
\right)\not e\left( \not k - \not k_2 \right)\gamma^\lambda v_2\bar
u_{B^\prime}\gamma^\rho\left( \not k + \not p_{B^\prime} \right)\gamma
^\nu u_B g_{\mu\nu}g_{\lambda\rho}}{\left[ (k+p_{B^\prime})^2
+i\delta \right]\left[ (k+p_A-k_2)^2+i\delta \right]\left[ (k-k_2)^2
+i\delta \right]\left[ k^2+i\delta \right]\left[ (k+q)^2+i\delta \right]}~.
\end{equation}
It was explained in details in Ref.~\cite{FM99}, that the
negative $t$-channel signature combination of the two pentagon
diagrams of Fig.~5, or, that is the same, the $s \leftrightarrow
- s$ antisymmetric part of the $D_2$, gets the contribution only
from the integration region where
\begin{equation}\label{a2}
\vec q^{~2} \sim \vec k^{~2} \sim |kp_1| \ll |kp_2| \sim s~,
\end{equation}
and therefore we can replace from the beginning
\begin{equation}\label{a3}
\frac{1}{(k+p_{B^\prime})^2+i\delta} \rightarrow P\frac{1}{2kp_2}
\end{equation}
and understand this singularity in a sense of the principal value
everywhere below. In order to simplify also the numerator of the
$D_2$ we use the familiar replacement (\ref{z}) for the
$t$-channel gluon propagators. These simplifications lead to the
following representation for $D_2$
$$
D_2 = -2ieq_fg^4s\bar u_{B^\prime}\frac{\not p_1}{s}u_B\int\frac
{d^Dk}{(2\pi)^D}\times
$$
$$
\frac{1}{\left[ (k+p_A-k_2)^2+i\delta \right]\left[ (k-k_2)^2+i\delta
\right]\left[ k^2+i\delta \right]\left[ (k+q)^2+i\delta \right]}
$$
\begin{equation}\label{a5}
\times\bar u_1\left\{ \not e\left( \not k - \not k_2 \right) +
\left( \not k + \not q \right) \not e + \frac{s}{2kp_2}\left(
x_1\not e\left( \not k - \not k_2 \right) - x_2\left( \not k +
\not q \right)\not e \right) \right\}\frac{\not p_2}{s}v_2~.
\end{equation}

Since the structure of $D_2$ is like the box diagram with two
massive external lines in opposite corners, it is convenient to
perform the Feynman parametrization joining first the pairs of
propagators which meet in each of the vertices with massless
external lines, and then to join two denominators obtained in
this way using the third Feynman parameter. Doing so we naturally
get the result of momentum integration in the form where the
dependence on the third Feynman parameter factorizes and the
integration over this parameter can be performed
straightforwardly. So we use in  Eq.~(\ref{a5}) the
representation
$$
\frac{1}{\left[ (k+p_A-k_2)^2+i\delta \right]\left[ (k-k_2)^2+i\delta
\right]\left[ k^2+i\delta \right]\left[ (k+q)^2+i\delta \right]} =
$$
\begin{equation}\label{a6}
6\int_0^1\frac{d^3yy_3(1-y_3)}{\left[ k^2 + 2k\left( (1-y_3)(q+y_1k_1)
- y_3y_2k_2 \right) + (1-y_3)\left( (1-y_1)t + y_1t_2 \right) + i\delta
\right]^4}~.
\end{equation}
For the term with $1/(2kp_2)$ in  Eq.~(\ref{a5}) we use the
relation
\begin{equation}\label{a7}
\frac{1}{a^4b} = 4\int_0^\infty\frac{du}{(a+bu)^5}~,
\end{equation}
which leads, together with  Eq.~(\ref{a6}), to the expression
$$
D_2 = -12ieq_fg^4s\bar u_{B^\prime}\frac{\not p_1}{s}u_B\bar u_1
\int_0^1d^3yy_3(1-y_3)\int\frac{d^Dk}{(2\pi)^D}\left\{ \frac{\not
e\left( \not k - \not k_2 \right) + \left( \not k + \not q \right)
\not e}{\left[ (k+p)^2 - y_3(1-y_3)b^2 + i\delta \right]^4}\right.
$$
\begin{equation}\label{a8}
\left.+ 4\int_0^\infty du\frac{s\left( x_1\not e\left( \not k -
\not k_2 \right) - x_2\left( \not k + \not q \right)\not e
\right)}{\left[ (k+p+up_2)^2 - y_3(1-y_3)b^2 - su\left(
(1-y_3)y_1x_1 - y_2y_3x_2 \right) + i\delta \right]^5}
\right\}\frac{\not p_2}{s}v_2~,
\end{equation}
where the notations
\begin{equation}\label{a9}
p = (1-y_3)(q+y_1k_1) - y_2y_3k_2~,\ \ \ b^2 = (1-y_2)\left( - (1-y_1)t
- y_1t_2 \right) + y_2\left( - (1-y_1)t_1 + y_1Q^2 \right)
\end{equation}
have been introduced.

Now the $k$-integration and then the integration over $u$ are
immediate to perform. The result is:
$$
D_2 = g\bar u_{B^\prime}\frac{\not p_1}{s}u_B\frac{2s}{t}eq_fg^3
\frac{\Gamma(2-\epsilon)}{(4\pi)^{2+\epsilon}}t\bar u_1\int_0^1
\frac{d^2y}{\left( b^2 \right)^{2-\epsilon}}P\int_0^1\frac{dy_3}
{\left[ y_3(1-y_3) \right]^{1-\epsilon}}\times
$$
\begin{equation}\label{a10}
\left\{ \left( \not q - \not p \right)\not e - \not e\left( \not
k_2 + \not p \right) + \frac {x_2\left( \not q - \not p
\right)\not e + x_1\not e\left( \not k_2 + \not p
\right)}{x_1\left( (1-y_3)y_1 - zy_3y_2 \right)}
\right\}\frac{\not p_2}{s}v_2~,
\end{equation}
with
\begin{equation}\label{a11}
z = \frac{x_2}{x_1}~.
\end{equation}
After some simplifying algebra with the use of the fact that $D_2$
enters into our result for the vertex only in the antisymmetric
under the replacement $1 \leftrightarrow 2$ combination (see Eq.~
(\ref{32})) we come to the representation
$$
D_2 = 2g\bar u_{B^\prime}\frac{\not p_1}{s}u_B\frac{2s}{t}eq_fg^3
\frac{\Gamma(2-\epsilon)}{(4\pi)^{2+\epsilon}}
$$
$$
\times\bar u_1t\frac{\not
p_2}{s}\int_0^1\int_0^1\frac{dy_1dy_2}{\left[ (1-y_2)\left( -
(1-y_1)t - y_1t_2 \right) + y_2\left( - (1-y_1)t_1 + y_1Q^2
\right) \right]^{2-\epsilon}}\biggl[ \frac {1-y_1}{x_1}
$$
$$
\times\left( 2x_2({k_1}_\perp e_\perp) - \not e_\perp{\not
k_1}_\perp \right)P\int_0^1
\frac{dy_3y_3^{\epsilon-1}(1-y_3)^{\epsilon}}{\left( (1-y_3)y_1 -
zy_3y_2 \right)} + x_2(ep_1)
$$
\begin{equation}\label{a12}
\times P\int_0^1\frac{dy_3y_3^{\epsilon-1}(1-y_3)^{\epsilon-1}}
{\left( (1-y_3)y_1 - zy_3y_2 \right)} + \frac{2}{\epsilon}\left(
\left(ek_1\right)(1-y_1) - x_2(ep_1) \right) \biggr]v_2~,
\end{equation}
and, finally, the relations
\begin{equation}\label{a13}
P\int_0^1\frac{dy_3y_3^{\epsilon-1}(1-y_3)^{\epsilon}}{\left(
(1-y_3)y_1 - zy_3y_2 \right)} = \frac{1}{\epsilon}\left( 1 -
2\epsilon^2\psi^\prime(1) \right)z^{-\epsilon}y_1^{\epsilon-1}
y_2^{-\epsilon}
\end{equation}
and
\begin{equation}\label{a14}
P\int_0^1\frac{dy_3y_3^{\epsilon-1}(1-y_3)^{\epsilon-1}}{\left(
(1-y_3)y_1 - zy_3y_2 \right)} = \frac{1}{\epsilon}\left( 1 -
2\epsilon^2\psi^\prime(1) \right)z^{-\epsilon}\left( y_1^{\epsilon-1}
y_2^{-\epsilon} - z^{2\epsilon-1}y_2^{\epsilon-1}y_1^{-\epsilon} \right)~,
\end{equation}
which are valid with an enough for us accuracy in the
$\epsilon$-expansion, lead to the result (\ref{34}).

\end{document}